\documentclass[pra,aps,singlecolumn,10pt,superscriptaddress,notitlepage,longbibliography,nofootinbib]{revtex4-2}
\usepackage[dvipsnames]{xcolor}
\usepackage[english]{babel} 
\usepackage{graphicx}
\usepackage{float}
\usepackage{braket}
\usepackage{epstopdf}
\usepackage{mathtools}
\usepackage{amsmath}
\usepackage{amssymb}
\usepackage{bbm,bm}
\usepackage[caption=false]{subfig}
\usepackage{pythonhighlight}
\usepackage{hyperref}
\usepackage{float}
\usepackage{bbold}
\usepackage[T1]{fontenc}

\usepackage{mathtools}
\usepackage{dsfont}
\usepackage{bm}
\usepackage{tikz}
\usetikzlibrary{external}
\graphicspath{{./Figures/}{./Figures_thesis/}}

\usepackage[markup=underlined]{changes}
\makeatletter
\@namedef{Changes@AuthorColor}{magenta}
\colorlet{Changes@Color}{magenta}
\makeatother
\usepackage{hyperref}
 \hypersetup{
     colorlinks=true,
     linkcolor=blue,
     filecolor=blue,
     citecolor = magenta,      
     urlcolor=red,
     }

\usepackage{braket}
\usepackage{xargs}
\newcommandx{\greencom}[2][1=]
{\todo[inline, color=green!40,#1]{#2}}
\newcommandx{\bluecom}[2][1=]
{\todo[inline, color=blue!40,#1]{#2}}
\newcommandx{\bluemargin}[2][1=]
{\todo[color=blue!40,#1]{#2}}

\usepackage{letltxmacro}
\LetLtxMacro{\ORIGselectlanguage}{\selectlanguage}
\makeatletter
\DeclareRobustCommand{\selectlanguage}[1]{%
  \@ifundefined{alias@\string#1}
    {\ORIGselectlanguage{#1}}
    {\begingroup\edef\x{\endgroup
       \noexpand\ORIGselectlanguage{\@nameuse{alias@#1}}}\x}%
}
\newcommand{\definelanguagealias}[2]{%
  \@namedef{alias@#1}{#2}%
}

\makeatother

\definelanguagealias{en}{english}
\definelanguagealias{EN}{english}
\definelanguagealias{eng}{english}
\definelanguagealias{de}{ngerman}

\newcommand{\ii}{\mathrm{i}}

\newcommand{\lp}{\left(}
\newcommand{\rp}{\right)}
\newcommand{\lsb}{\left[}
\newcommand{\rsb}{\right]}

\usepackage{setspace}
\usepackage{xr}
\makeatletter

\newcommand*{\addFileDependency}[1]{% argument=file name and extension
\typeout{(#1)}% latexmk will find this if $recorder=0
\@addtofilelist{#1}
\IfFileExists{#1}{}{\typeout{No file #1.}}
}\makeatother

\begin{document}

\title{Direct space-time modeling of mechanically dressed dipole-dipole interactions with electromagnetically-coupled 
oscillating dipoles}

\author{Yi-Ming Chang}
\email[]{yiming.chang@queensu.ca}
\affiliation{Department of Physics, Engineering Physics and Astronomy, Queen's University, Kingston ON K7L 3N6, Canada}
\author{Kamran Akbari}
%\email[]{shughes@queensu.ca}
\affiliation{Department of Physics, Engineering Physics and Astronomy, Queen's University, Kingston ON K7L 3N6, Canada}
%------------------------------------------------------------------
\author{Matthew Filipovich}
%\email[]{shughes@queensu.ca}
\affiliation{Department of Physics, 
%Clarendon Laboratory,
University of Oxford,
Parks Road Oxford
OX1 3PU, United Kingdom}

\author{Stephen Hughes}
\email[]{shughes@queensu.ca}
\affiliation{Department of Physics, Engineering Physics and Astronomy, Queen's University, Kingston ON K7L 3N6, Canada}
%------------------------------------------------------------------
\date{\today}

\begin{abstract} 

We study the radiative dynamics of coupled electric dipoles, modelled as Lorentz oscillators (LOs), in the presence of real-time mechanical oscillations. The dipoles are treated in a self-consistent way through a direct electromagnetic simulation approach that fully includes the dynamical movement of the charges, accounting for radiation reaction, emission and absorption. This allows for a powerful numerical solution of optomechanical resonances without any perturbative approximations for the mechanical motion. 
The scaled population (excitation) dynamics of the LOs are investigated as well as the emitted radiation and electromagnetic spectra, which demonstrates how the usual dipole-dipole resonances couple to the underlying Floquet states, yielding multiple spectral peaks that are separated from the superradiant and subradiant states by an integer number of the mechanical oscillation frequency. 
Moreover, we observe that when the mechanical amplitude and frequency are sufficiently large, these additional spectral peaks undergo further modification, including spectral splitting, spectral squeezing, or shifting. These observations are fully corroborated by a theoretical Floquet analysis conducted on two coupled
harmonic oscillators.

\end{abstract}

\maketitle
\section{Introduction}

The spontaneous emission (SE) process can be modelled from the viewpoint of vacuum fluctuations in quantum field theory or from radiation reaction~\cite{Milonni_Semiclassical_1976}, with the latter mapping on to classical and semiclassical field-matter theories, whereby the radiative dynamics can  be captured through polarization dipoles and Lorentz oscillators (LOs)~\cite{Barnes_Classical_2020} coupled to classical fields. 

In both classical and quantum field pictures, the SE rate is governed by the (projected) photonic local density of states (LDOS)~\cite{Gerry_Introductory_2005,Novotny_Principles_2012,Joulain_Definition_2003,Forati_Spontaneous_2022}, which is related to the imaginary part of the system's electric field Green function $\mathbf{G}$, through
\begin{equation}\label{eq: SE_rate_LDOS}
\gamma({\bf r}_0,\omega_0) = \frac{2{\bf d} \cdot {\rm Im} {\bf G}({\bf r}_0,{\bf r}_0,\omega_0) \cdot {\bf d}}{\epsilon_0\hbar},
\end{equation}
for an emitter at position ${\bf r}_0$, with dipole moment ${\bf d}$ (assumed real) and resonance frequency $\omega_0$. The classical Green function of the medium is defined from
\begin{equation}
%\kappa_b(\mathbf{r},\omega)
\boldsymbol{\nabla}\times\boldsymbol{\nabla}\times\mathbf{G}(\mathbf{r},\mathbf{r}',\omega) -\frac{\omega^2}{c^2}\epsilon(\mathbf{r},\omega)\mathbf{G}(\mathbf{r},\mathbf{r}',\omega)
=\frac{\omega^2}{c^2}\mathbf{I}\delta(\mathbf{r}-\mathbf{r}'),
\label{eq: GFHelmholtz2}
\end{equation}
where  
%$\kappa_b(\mathbf{r},\omega)=\mu_b^{-1}(\mathbf{r},\omega)$, 
$\epsilon({\bf r},\omega)$ is the dielectric constant of the medium, $c$ is the speed of light in vacuum, and ${\mathbf{I}}$ is the unit tensor (dyad), and we assume a nonmagnetic medium. 

\begin{figure}[t]
    \centering
    % \scalebox{.99}{\input{Schematics_Overview}}
    \includegraphics[width=0.6\linewidth]{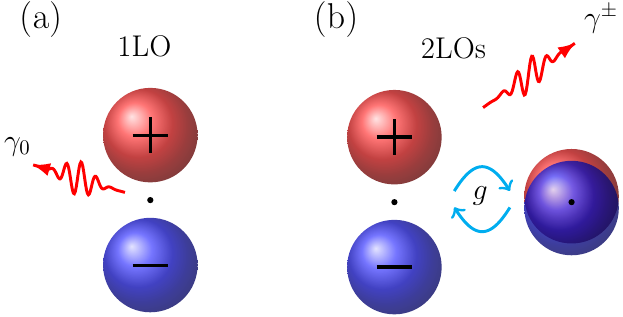}
    \caption{Schematic diagrams of (a) a single LO initially in its excited state located in free space having the decay rate $\gamma_0$ defined in Eq.~\eqref{eq: free space decay rate}, and (b) two coupled LOs with coupling rate $g\equiv g_{12}$ given in Eq.~\eqref{g_ij}, one initially in the excited state and the other in the ground state, yielding subradiant and superradiant radiative decay processes with rates $\gamma^{\pm}=\gamma_0\pm\gamma_{12}=g_{21}$, with $\gamma_{12}$ defined in Eq.~\eqref{gamma_12}.
    }
    \label{fig: schematics_1LOs_2LOs}
\end{figure}

For example, for a dipole in free space, as depicted in Fig.~\ref{fig: schematics_1LOs_2LOs}(a), we have
${\rm Im}{G}_{ii}=\omega^3/(6\pi c^3)$, and % $\mathbf{n}_b = 1$
\begin{equation}\label{eq: free space decay rate}
\gamma_0(\omega_0) = \frac{d^2 \omega_0^3}{3\pi \epsilon_0 \hbar c^3},
\end{equation}
 is the free space SE decay rate,
for an emitter with resonance frequency $\omega_0$
with  $d=\lVert{\bf d}\rVert$.
In a dielectric with
$\epsilon_{\rm B}= n_{\rm B}^2$, this rate is simply modified by $n_{\rm B}$, the background refractive index of the medium.
Moreover, for any inhomogeneous 
structure, the SE rate can be controlled by the LDOS (or projected LDOS, as the dipole direction also matters), which is modified by the electromagnetic environment surrounding the emitters. The pioneering work of Edward Purcell first pointed out that the SE rate of quantum emitters could be altered by placing them inside a resonant cavity~\cite{Purcell_Resonance_1946}, causing enhancement or inhibition of the SE rate. Subsequent research has extensively explored this process across various physical systems, such as emitters within photonic crystals~\cite{Ogawa_Control_2004,Lodahl_Controlling_2004,Angelakis_Photonic_2004,Dignam_Spontaneous_2006,Hughes_Coupled-cavity_2007,Noda_Spontaneous-emission_2007,Lodahl_Interfacing_2015}.

An alternative approach for modifying the SE dynamics involves coupling to other atoms, or to a collection of atoms, typically in the
near-field regime (i.e., much less than their emission wavelength). This results in  collective excitations that decay either faster (superradiance) or slower (subradiance) than the individual emitters, 
which has been demonstrated in a number of systems~\cite{DeVoe_Observation_1996,Eschner_Light_2001,Mokhlespour_Collective_2012,VanVlack_Spontaneous_2012,Trebbia_Tailoring_2022}. These coupled dipole studies reveal that the modified SE rate is dependent on interatomic distances, which can significantly impact the amplitude and width of the emission spectrum~\cite{Mokhlespour_Collective_2012,Holzinger_Nanoscale_2021}.

For two atoms, with their center-of-mass (COM) at positions ${\bf R}_n$ and with  resonant energies $\omega_n$, with $n=1,2$,
 then we also have an incoherent photon transfer rate
\cite{Angelatos_Entanglement_2015}\footnote{Here we also allow for complex dipole moments}
\begin{equation} % \label{eq: gamma_12}
\gamma_{12} = \frac{2{\bf d}_1 \cdot {\rm Im} {\bf G}({\bf R}_1,{\bf R}_{2},\omega_{2}) \cdot {\bf d}_{2}}{\epsilon_0 \hbar},
\label{gamma_12}
\end{equation}
and a coherent coupling rate
\begin{equation} % \label{eq: frequency_shift}
g_{12} = -\frac{{\bf d}_{1} \cdot {\rm Re} {\bf G}({\bf R}_1,{\bf R}_{2},\omega_{2}) \cdot {\bf d}_{2}}{\epsilon_0 \hbar},
\label{g_ij}
\end{equation}
where a Markov approximation has been applied (e.g., the rates are assumed to be constant over the frequency range of interest, so all rates are instantaneous in time).
% Once again, both of these effects can be derived quantum mechanically~\cite{Dung_Resonant_2002} or classically~\cite{Novotny_Principles_2012}.

Note also that the above expressions
can be derived classically or quantum mechanically, with a full classical-quantum correspondence
(discussed in more detail in Appendix \ref{sec:dip-dip}). 
%, and the Lamb shift determines the coupling $g_{nn'}$ between the atoms in both approaches. 
For example, in the quantum case, when the dipole bosonic operators $b_n$ and $b_n^\dagger$ are introduced for each atom, the  interaction Hamiltonian for dipole-dipole coupling is  $H_{dd}=\hbar g_{12}(b_{1}+b^\dagger_{1})(b_{2}+b^\dagger_{2})$, and within a rotating-wave approximation,
$H_{dd}^{\rm RWA}=\hbar g_{12}(b_{1}^\dagger b_2+b^\dagger_{2} b_1)$. 
With the appropriate initial condition, this can lead to the creation of superradiant states (Dicke states~\cite{Dicke_Coherence_1954}, specialized to two dipoles), which decay with the rate $\gamma^+=\gamma_0+\gamma_{12}$, or subradiant states, which decay at the rate $\gamma^-=\gamma_0-\gamma_{12}$. For simplicity, here we assume equal emitter strengths and resonance frequencies, so that $\gamma_0 = \gamma_{11}=\gamma_{22}$, $\gamma_{12}=\gamma_{21}$, and $g_{12}=g_{21}\equiv g$ [see Fig.~\ref{fig: schematics_1LOs_2LOs}(b)].

Recently, the direct numerical  modelling of dipole-dipole interactions, with no intrinsic approximations (such as the dipole approximation, rotating wave approximation or Markov approximation), was demonstrated by directly solving the self-consistent electrodynamics simulations of LOs and oscillating point charges~\cite{Filipovich_PyCharge_2022}. This direct modelling approach can have enormous benefits, as it 
%powerful 
connects classical dynamical simulations to complex SE dynamics, often in ways that are intractable or difficult to account for in standard quantum theory (or standard classical theory). For example, 
%in solving Maxwell's equations, 
one does not have to make a rotating wave approximation, and one can also go beyond a dipole approximation, two common approximations in quantized light-matter interaction theories.  

A particularly intriguing way of 
modifying the SE dynamics is to move the COM of the emission dipole itself, e.g., modeled as a moving atom~\cite{Dalibard_Fundamental_1992}. At relativistic velocities, this can yield very small corrections to $\gamma_0$~\cite{Boussiakou_Quantum_2002,Cresser_Rate_2003}. A more practical and significant way to modify the
SE rate is through mechanical perturbations of the photonic environment, resulting in an optomechanical coupling, e.g., where the SE can be altered by a vibrating cavity wall~\cite{Aspelmeyer_Cavity_2014Book,Aspelmeyer_Cavity_2014,Kippenberg_Cavity_2007}. As is well known, if the atom is close to a mirror (e.g., a half-open cavity), the emitted photon can be reflected by the mirror, which (depending on the angle of emission, frequency, and precise distance from the mirror) can either create a constructive or destructive interference at the position of the atom~\cite{Novotny_Principles_2012,Dorner_Laser-driven_2002,Eschner_Light_2001,Wilson_Vacuum-field_2003,Beige_Spontaneous_2002}.
 
Indeed, it is now well known that the SE rate and the effective level (frequency) shift of a quantum emitter, such as a two-level atom, can be modified by its interaction with an oscillating mirror. This interaction manifests %as a modulation of 
in a  positional-dependent SE rate and effective level shift, scaling with $\cos(2k_0R_0)$ and $\sin(2k_0R_0)$, respectively, where $k_0$ is the wave vector and $R_0$ is the atom-mirror separation~\cite{Glaetzle_Single_2010,Ferreri_Spontaneous_2019}. In 
Ref.~\cite{Glaetzle_Single_2010}, 
using a rotating wave approximation and one-dimensional quantized fields,
the SE of a two-level system 
%(TLS) 
near an oscillating plate was investigated. The study focused on field modes propagating within a limited solid angle, $\theta,$ towards the mirror and  reflected back to the atom, influenced by the mirror oscillation amplitude. 
In a Markov approximation (instantaneous coupling), when $\gamma\tau \ll 1$, $v_{\rm mirror} = R_{M}\omega_{M} \ll c$ and $\omega_{M}\tau\ll 1$ (with $\tau=2R_0/c$ the round-trip time, and $R_{M}$ and $\omega_{M}$ are the (mechanically) oscillating amplitude and frequency of the mirror), the modified SE rate 
was found to be
\begin{equation}\label{eq: modified gamma}
\gamma' = \gamma[1-\theta {J}_0(2k_0 R_{M}) \cos(2k_0R_0)],
\end{equation}
and similarly for the atom frequency,
%\begin{equation}\label{eq: modified frequency}
$\omega_0' = \omega_0 - \theta \frac{\gamma}{2} {J}_0(2k_0 R_M) \sin(2k_0R_0)$,
%\end{equation}
where $\theta$ is less than 1 and depends on the scattering geometry, 
%$k_0=\omega_0/c$ is the wavenumber, 
and ${J}_0$ is the zeroth-order Bessel function of the first kind, which represents the mirror's oscillation effect. Modulating the mirror with the dependence $R_M\sin(\omega_M t)$, yields the following amplitude for an initially excited atom:
\begin{align}\label{eq: population amplitude}
C(t) &= e^{-\ii\omega_0' t}
e^{-\gamma' t/2} 
%\nonumber \\
%& 
\exp\left [
-\theta\frac{\gamma}{2} e^{\ii\omega_0 \tau}
\sum_{n \neq 0}
{J}_n(2k_0 R_M) \frac{e^{-\ii n \omega_M t}-1}{\ii n\omega_M}
\right],
\end{align}
where $|C(t)|^2$ is the time-dependent population.
 The excited atom population decays exponentially at a modified rate, $\gamma'$, which includes an oscillating modulation, where the amplitude of this modulation decreases when the oscillation frequency of the mirror increases.
In the resolved sideband regime, where $\gamma<\omega_M$, this modification is small. 

In Ref.~\cite{Ferreri_Spontaneous_2019}, the authors considered a more general model, where a 
full three-dimensional electromagnetic field is quantized with the time-dependent boundary conditions determined by the (adiabatically) oscillating mirror. A general orientation of the atomic dipole moment was considered, but the oscillating frequency is only a few GHz compared with the transition frequency of an atom ($\approx1$ THz for that work), and the oscillating amplitude was $R_M/R_0=0.1$.
Another work investigated the SE rate and emission spectrum of a TLS positioned within a dynamic photonic crystal, which showed how SE modifications relate to the time-dependent photonic density states~\cite{Calajo_Control_2017}. These findings suggested that the dynamical control of the boundary conditions causes a further modification of the SE process~\cite{Glaetzle_Single_2010,Calajo_Control_2017,Ferreri_Spontaneous_2019}. This is a nontrivial regime as the SE process is now non-stationary because of the time-dependent boundary conditions. 

Apart from the interest in fundamental optics,
the investigation of vibrational atoms and molecules is also important in the context of Raman spectroscopy and Surface-Enhanced Raman Spectroscopy (SERS), which forms a model of molecular optomechanics~\cite{Roelli_Molecular_2015,Dezfouli_Molecular_2019,Esteban_Molecular_2022}. Molecular optomechanics provides insights into the fundamental principles governing the interaction between light and matter, as well as the development of technologies or applications, such as ultrasensitive detectors that can measure tiny forces and displacements at atomic scale~\cite{Esteban_Molecular_2022,Patra_Molecular_2023,Batignani_Accessing_2020}.

\begin{figure}[ht]
    \centering
    % \scalebox{.80}{\input{Figure1}}
    \includegraphics[width=0.85\linewidth]{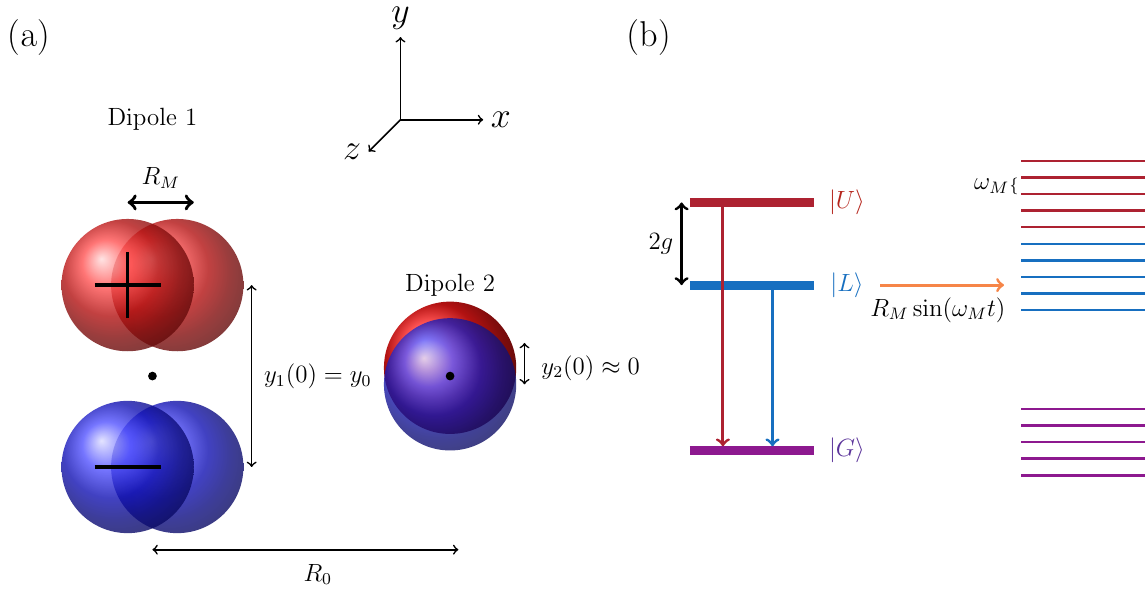}
    \caption{(a) Schematic diagram of two coupled LOs with mechanical oscillation applied on the origin of Dipole 1 (along the $x$ direction), and its energy level diagram with and without mechanical drive, where Dipole 1 is initially excited. Here, $R_M$ is the mechanical oscillating amplitude of Dipole 1. The distance $R_0$ is the initial dipole-dipole separation along the $x$-axis, $y_i(0)$ are the initial separations between the opposite charges along $y$-axis at time $t=0$. We will consider $y_0 = 1\,$nm, as the maximum charge oscillating amplitude. The black dot represents the origin (COM) of the dipoles. 
    (b) Schematic diagram of the energy levels for the coupled LOs without (left) and with (right) mechanical motion, where the joint two-atom-dressed ground state $\lvert G\rangle$, lower polariton state $\lvert L\rangle$ and upper polariton state $\lvert U\rangle$ are shown. 
    After turning on the mechanical drive, the states lead to manifolds of dressed states, 
    where the sub-states are separated by an integer of the driving frequency $\omega_M$. 
    }
    \label{fig: schematics}
\end{figure}

In this paper, we introduce
a direct space-time model for
coupling mechanical oscillations
to radiatively coupled emitters, 
where we  directly move (oscillate) the COM of the  dipoles mechanically, e.g., instead of vibrating mirrors which cause phonons to couple with photons~\cite{Glaetzle_Single_2010,Ferreri_Spontaneous_2019}. 
The main numerical simulations are carried out using PyCharge,
a recently developed Python package that can calculate the electromagnetic fields and (vector and scalar) potentials generated by moving point charges and can self-consistently simulate dipoles as 
LOs~\cite{Filipovich_PyCharge_2022}. The electric dipoles (LOs) are formed by a coupled oscillating point charges
with opposite signs, equal magnitude and a separated displacement [see Fig.~\ref{fig: schematics_1LOs_2LOs}(a)].
The charges oscillate around the COM of the electric dipole along the direction of polarization. The origin of the LOs can be set to stationary [see Fig.~\ref{fig: schematics_1LOs_2LOs}(b)] or move as a function of time [see Fig.~\ref{fig: schematics}(a)]. The LOs are naturally damped within the self-consistent solver, with radiative decay. The motion of the LOs is restricted to nonrelativistic velocities; however,  PyCharge can model a wide range of effects 
in electromagnetism, including  ``Braking radiation'' (Bremsstrahlung)~\cite{Filipovich_Space-time_2021}.

There are several benefits to using PyCharge to model the dynamics of LOs. First, it has already been shown to be very accurate for obtaining the known rates of dipole-dipole coupling from QED theory (typical numbers are $<0.2\%$ relative error, even for very close dipole separations, deep sub-wavelength)~\cite{Filipovich_PyCharge_2022}.
Secondly, PyCharge can explore effects that cannot be solved analytically or via the usual simulation software for electromagnetism. For instance, the numerical simulation of PyCharge does not rely on a Markovian approximation, dipole approximation, nor any rotating wave approximation, which are often required in other approaches, such as quantum master equations.

Specifically, in this work we use PyCharge to explore the modified SE decay dynamics from a pair of atoms (treated as $s$-polarized dipoles) that are coupled through photon-coupling effects such as F\"orster coupling~\cite{Forster_Zwischenmolekulare_1948}, when one of the atoms are subject to a mechanical oscillation on the COM. In general, an atom subject to a periodic oscillation will form Floquet states~\cite{Bukov_Universal_2015,Oka_Floquet_2019,Takashi_Floquet_2023}, which can then {\it dress} the usual dipole-dipole-induced superradiant and subradiant states to have multiple resonances, separated (in the perturbative limit) by $\pm l\omega_M$ ($l$ integer). 
Thus, 
 we also develop a rigorous Floquet analysis of two-coupled atoms (harmonic oscillators)  with mechanical perturbations, to help with the interpretation and prediction of our direct
 time-dependent solution using PyCharge. 
In PyCharge, the dipoles are modeled as 
 classical LOs~\cite{Novotny_Principles_2012}, by directly solving the equation of motion for the LOs including the scattered fields responsible for modified SE. 

The rest of our article is organized as follows: In Sec.~\ref{sec: theory} we present the main theoretical background for the direct
electromagnetic simulations in PyCharge. 
In Sec.~\ref{sec: theory2},
we then describe an analytical model
of two coupled quantum harmonic oscillators (HOs),
with periodic modulation the dipole-dipole coupling, and show how this can be solved using
Floquet theory. 
In Sec.~\ref{sec: results}, we present
several example results showing the electric dipole dynamics of two coupled LOs with and without mechanical oscillations. 
We also show results for the emitted spectra,  
for different driving parameters, 
and each 
example regime is corroborated using the analytical Floquet theory. 
Finally, in Sec.~\ref{sec: conclusion}, we give a summary and conclusions.
In addition, we present two Appendices.
In Appendix \ref{sec:dipole-fields},
we present the field solutions 
from oscillating point dipole
in a homogeneous medium.
While in
Appendix \ref{sec:dip-dip}, 
we  present a 
quantum and classical
field theory
for dipole-dipole coupling,
which we use in the main text
to derive a Floquet solution with periodic 
oscillations in the dipole-dipole coupling. 

\section{Two Oscillator Optomechanics: Theory and numerical implementation of the Direct Electromagnetic Simulation}\label{sec: theory}

To numerically model the radiative decay through 
radiation reaction, we start with one initially excited dipole. Figure~\ref{fig: schematics} shows a schematic of the two coupled LOs with mechanical oscillations governed by Eq.~\eqref{eq: 2LOs Mec}, as well as the potential energy levels. The LOs are formed by two opposite point charges, $\pm q_n$ and masses $m_{n\pm}$
% , respectively, so that the effective mass is $m_{{\rm eff},n}=m_{n+}m_{n-}/(m_{n+}+m_{n-})$) 
for each dipole [having the effective (reduced) mass $m_{{\rm eff},n}=m_{n+}m_{n-}/(m_{n+}+m_{n-})$], with a separate displacement, which oscillates around the origin along the $y$-axis ($s$-polarized dipoles). 
We can then set an initial displacement for the dipoles, from $y_n(t=0)$. For example, we can set $y_1(t=0)=y_0$ and $y_2(t=0)\approx{0}$, where $y_0=1\,$nm. Thus, we 
%can 
consider Dipole 1 is initially excited, and Dipole 2 will become excited through interactions from the electric field generated by Dipole 1. As they oscillate, the LOs will radiate (dissipate) energy that causes the
dipole moment to decrease
\cite{Novotny_Principles_2012}, but potentially oscillate when coupled to nearby LOs.

Considering two spatially separated
%(internally two-body) 
LOs with the same effective mass, $m_{{\rm eff},1}=m_{{\rm eff},2}=m_{{\rm eff}}$, the same charge, $q_1=q_2=q$, and the same transition frequency,  $\omega_1=\omega_2\equiv\omega_0$, in free space,   the coupled dipole equations of motion are
\begin{subequations}\label{eq: 2LOs Mec}
\begin{align}
&\ddot{\mathbf{d}}_1+\gamma_0\dot{\mathbf{d}}_1+\omega_0^2\mathbf{d}_1=\frac{q^2}{m_{\rm eff}}\mathbf{E}_{d}(\mathbf{R}_1(t),t),\label{eq16a}\\
&\ddot{\mathbf{d}}_2+\gamma_0\dot{\mathbf{d}}_2+\omega_0^2\mathbf{d}_2=\frac{q^2}{m_{\rm eff}}\mathbf{E}_{d}(\mathbf{R}_2,t),\label{eq16b}
\end{align}
\end{subequations}
where ${\bf E}_d$ is the component of the total electric field in
the direction of polarization generated by all sources in the system (including scattered fields), which is the sum of Coulomb and radiative electric field contributions,
%\ymc{, exclude its own charges}, %explicitly 
given in Eqs.~\eqref{eq: E_coul} and~\eqref{eq: E_rad} of Appendix.~\ref{sec:dipole-fields}, respectively. 
In the coupled dipole equation of motion,  $\mathbf{R}_1(t)$ is the position vector of the COM of Dipole 1 which is in periodically mechanical motion, ${\bf R}_2$ (assumed fixed) is the position vector of the COM of Dipole 2.
Since any relevant physics here depends on the relative distance, we define ${\bf R}(t)={\bf R}_1(t)-{\bf R}_2$. Thus, the mechanical motion is defined from
\begin{equation}\label{eq: mechanical motion}
    \mathbf{R}(t) = \mathbf{R}_0+\mathbf{R}_M\sin(\omega_{M}t),
\end{equation}
where
%, without loss of generality, 
we let ${\bf R}_2={\bf 0}$, ${\bf R}_0=R_0{\bf \hat{x}}$ and ${\bf R}_M=R_M{\bf \hat{x}}$, so that $R_M$ is the mechanical oscillating amplitude of Dipole 1, $R_0$ is the initial static distance between the COMs of the two dipoles along $x$-axis, and  $\omega_{M} = 2\pi/T$ is the mechanical oscillating frequency.
Therefore, the relative time-dependent distance between the two LOs is simply $R(t)\equiv\lVert{\bf R}(t)\rVert=R_0+R_M\sin(\omega_Mt)$ [see Fig.~\ref{fig: schematics}(a)].

For our study, 
we will assume that 
the average  mechanical velocity of the dipole is $v_{M} \approx R_M\omega_{M} \ll c$. In addition, we constrain the movement of dipoles and charges to non-relativistic velocities, so that the dissipation of charges due to radiation reaction damping is negligible. Thus, we set a reasonable maximum velocity of charge to be $c/100$. 
Thus, the mechanical oscillation parameters $R_M$ and $\omega_{M}$ play  the significant role in our study, and we primarily focus on their respective dimensionless ratios, denoted as
${R_M}/{R_0}$ which is constrained to values less than $0.8$ (this limitation is imposed to ensure that the dipoles remain spatially separate and do not come in direct contact with each other)
and
${\omega_{M}}/{g}$.

In the SE rate regime,
the energy decay rate  of an emitted photon is equivalent to the population decay rate, where the free-space decay rate, $\gamma_0$,
can also be modelled as 
an Einstein A coefficient~\cite{Milonni_Coherence_2010}.
However, it is important to emphasize that the relationship between the dynamics of LOs and the population of two-level systems' states can only be connected in a weak excitation regime (since the LO model does not account for a two-level system saturation effects). 

To connect to the LO equivalent of
population decay,
one can also calculate the self-consistent dipole moment of the LOs at each time step by solving Eq.~\eqref{eq: 2LOs Mec}. 
The total energy $\mathcal{E}$ of an oscillating dipole, is then given by 
\begin{equation}\label{eq: total energy}
    \mathcal{E}_n(t)=\frac{\omega_0^2m_{{\rm eff}}}{2q^2}d_n^2(t)+\frac{m_{{\rm eff}}}{2q^2}\lsb\frac{\mathrm{d}}{\mathrm{d}t}{d}_n(t)\rsb^2,
\end{equation}
which is the sum of the potential energy and kinetic energy of the dipole, respectively, and 
$d_n(t)$ is the time-dependent dipole moment of the $n$th dipole. We assume that the total energy $\mathcal{E}_n$ is proportional to the population of the excited state (mapping on to a two-level system model), due to the equivalence of the energy decay rate and the population decay rate. %The probability of excitation means the probability for electrons in an atom occupies in excited states.
Subsequently, we can write the scaled population of the excited states as the normalized total energy~\cite{Filipovich_PyCharge_2022},
\begin{equation}\label{eq: scaled population}
    \widetilde{\mathcal{N}}_n^{{\rm e}}(t)=\frac{\mathcal{E}_n(t)}{{\rm max}(\mathcal{E}_n)}.
\end{equation}
for the $n$th dipole.

\section{Floquet theory of two coupled oscillators with
periodic coupling}\label{sec: theory2}

As described in Appendix.~\ref{sec:dip-dip},
the system Hamiltonian for two coupled {\it quantum} 
HOs, when considering the coupling between two electric dipoles
in free space, 
with equal oscillator frequencies
$\omega_0$, takes on the form
\begin{equation}
    \begin{split}
        H = \hbar\omega_0 b_1^\dagger b_1 + \hbar\omega_0 b_2^\dagger b_2 + \hbar g(b_1+b_1^\dagger)(b_2+b_2^\dagger),
    \end{split}
    \label{H_qm}
\end{equation}
where $b_n$ ($b_n^\dagger$) is the bosonic annihilation (creation) operator of the $n$th dipole and we 
assume the dipoles are coupled through 
a near-field electromagnetic interaction,
with $g =q_1 q_2 /(8 \pi \hbar \epsilon_0m_{\rm eff}\omega_0R^3)$,
where 
$R =\lVert{\bf R}_1 - {\bf R}_2\rVert.$
More generally, this
interaction strength
can be obtained from
Eq.~\eqref{g_ij}, which depends on the real part of the Green function, ${\bf G}({\bf R}_1,{\bf R}_2)$.
A correspondence between the 
%fully
classical dipole model can easily be made by 
connecting
$q_n^2/m_{{\rm eff},n} = 2 \omega_n d_n^2/\hbar$ (see Appendix.~\ref{sec:dip-dip}, and also Ref.~\cite{Filipovich_PyCharge_2022}),
and in terms a quantum dipole moment, one can also show that $g=g_{12}$, introduced earlier
(see also Refs.~\cite{Wubs_Multipole_2003,Kristensen_Decay_2011}).

Below we will use the  Hamiltonian for the  quantum 
HO problem, but we stress that exactly the same resonances occur for a classical Hamiltonian system, which is easy to recover by simply replacing the harmonic oscillator operators by (classical) $c$-numbers. This classical correspondence is possible since the system obeys a linear response, where the resonances yield the polariton frequencies.
Interestingly, note that the specific form of the system Hamiltonian is similar to a
quantum Hopfield-like model without the 
diamagnetic term~\cite{Muniain_Description_2025,Hughes_Reconciling_2024}, which is needed for describing the coupling between
a quantum dipole and a single quantized cavity mode, otherwise 
a phase transition (often ill-defined) can occur when
$\eta=g/\omega_0>0.5$. In the present case, such a limit cannot be reached, since the dipole approximation
is only valid when
$R>4a$, where 
$a$ is the size (length) of the dipole (for example, the dipole moment
$d=e a$).

In order to understand the dynamics of the system and calculate the {\it normal modes} or resonances (in the absence of dissipation), we write the Heisenberg equation of motion as $\mathrm{d}\boldsymbol{B}/\mathrm{d}t=-\ii\mathbf{H}\,\boldsymbol{B}$, with  $\boldsymbol{B}=[b_1, b_2, b_1^\dagger, b_2^\dagger]^T$,
where 
\begin{equation}
    \begin{split}
        \mathbf{H} &
        =\hbar\omega_{0}\left[\begin{array}{cccc}
            1 & \eta & 0 & \eta 
            \\
             \eta & 1 & \eta & 0 
             \\
             0 & -\eta & -1 & -\eta 
             \\ 
             -\eta & 0 & -\eta & -1
        \end{array}\right],
    \end{split}
    \label{MatrixH_EoM}
\end{equation}
%where $\eta=g/\omega_{0}$. The 
and the normal modes of the hybrid system are $\omega_\pm=\sqrt{\omega_0^2\pm2g\omega_0}$, which are 
net positive 
%the stable upper/lower polaritons 
since $g< \omega_0/2$ for dipole-dipole interactions within the dipole approximation. 

We next present an analysis of the two coupled HOs with relative COM movement where the system is {\it Floquet engineered} through the time-dependent dipole-dipole coupling rate $g\to g(t)$. 
This time-dependent character of the system is through the time-dependent factor to the interaction Hamiltonian $H_{\rm dd}={d}_1{d}_2/(4\pi\epsilon_0 R^3)$ via 
$R=R_0\to R(t)=R_0+R_M\sin\omega_Mt$, which gives $g\to g(t)$,
where
\begin{equation}
g(t)= \frac{g}{[1+(R_M/R_0)\sin\omega_Mt]^3}.
\end{equation}
Due to the periodicity of the time-dependent coupling, with period $T$, the Hamiltonian is also periodic: ${H}(t) ={H}(t+T)$, 
and so is the dynamical matrix of the equation of motion: $\mathbf{H}(t) =\mathbf{H}(t+T)$.

We  can thus expand the time-dependent Hamiltonian as a Fourier series 
$\mathbf{H}(t)=\sum_{m\in\mathbb{Z}}\mathbf{H}_m\,\mathrm{e}^{\ii m\omega_{ M}t}$, with $\mathbf{H}_m=({1}/{T})\int_{T}\mathrm{d}t\,\mathrm{e}^{-\ii m\omega_Mt}\,\mathbf{H}(t)$, yielding
\begin{equation}
    \begin{split}
        \mathbf{H}_{m} &
        =%\omega_0\delta_{m0}\mathbf{I}_{2\otimes2}+
        \hbar\left[\begin{array}{cccc}
            \omega_{0}\delta_{m0} & g_m & 0 & g_m 
            \\
             g_m & \omega_{0}\delta_{m0} & g_m & 0 
             \\
             0 & -g_m & -\omega_{0}\delta_{m0} & -g_m 
             \\ 
             -g_m & 0 & -g_m & -\omega_{0}\delta_{m0}
        \end{array}\right],
    \end{split}
    \label{H_m}
\end{equation}
 % \begin{equation}
% \begin{split}\displaystyle
%  {H}_m
%  &=\lp\frac{\hbar\omega_0}{2}\sigma_{z,1}+\frac{\hbar\omega_0}{2}\sigma_{z,2}\rp\,\delta_{m0}+\hbar{g}_m\,\sigma_{x,1}\sigma_{x,2},
% %  \\
% % {H}_{m\neq0}&=\hbar{g}_m\sigma_{x,1}\sigma_{x,2}   
% % ,
% \end{split}
% \label{H_m}
% \end{equation}
where  
  \begin{equation}
\begin{split}\displaystyle
 % \hbar{g}_0
 % &=g_0^\mathrm{D}\lsb\frac{1}{T}\int_{-T/2}^{T/2}dt\,\frac{1}{\lp 1-\widetilde{R}_M\sin\omega_Mt\rp^3}\rsb
 % \\
 \hbar{g}_m
 &=
 %%\frac{d_1 d_2}{4\pi\epsilon_0\omega_0 R_0^3}
 \hbar g\left(\frac{1}{T}\int_{-T/2}^{T/2}\mathrm{d}t\,\frac{\mathrm{e}^{-\ii m\omega_Mt}}{[1+(R_M/R_0)\sin\omega_Mt]^3}\right)
.
\end{split}
\label{g_m}
\end{equation}

Notably, we see that the static coupling rate, $g_0$, is renormalized by the average of the time-dependent term~\cite{Akbari_Floquet_2025}:  
\begin{equation}
\hbar g_0=\hbar g\,\frac{1+0.5(R_M/R_0)^2}{[1-(R_M/R_0)^2]^{2.5}}\geq\hbar g,
\end{equation}
and thus the mechanical drive renormalizes the dc Hamiltonian through an effective increase in 
$g$, which is the undriven coupling term.
Later, we will see such an effect in our simulations 
when $R_M$ is sufficiently large.

% (later, in our simulations, one can see this effect, e.g., by comparing the transparent thick dashed and transparent thick solid lines in the eigenenergy  bases of panel (c)
% of Figs.~\ref{fig: PyCharge simulation 1}, \ref{fig: PyCharge simulation 2} and \ref{fig: PyCharge simulation 3}). 
% Note this DC-renormalization (comparing the thick dashed lines with the thick solid eigenenergy lines) may show negligible difference as the static coupling $g$ is small in this study, and because $g$ is very small, in order to see big effect we need large enough $R_M$.
% This is due to the fact that the mechanical oscillation introduces the time-dependence through a nonlinearity, here $1/R^3$, similar to the Floquet engineering problem~\cite{Akbari_Floquet_2025}, however, here it almost coincides with the Stark shift that is the splitting and deflection of the quasienergies from their originating eigenenergies as a drive parameter changes~\cite{Stark_Beobachtungen_1914}, due to smallness of $g$.
% When
% $R_{M} \rightarrow 0$, then
% we recover the time-independent  Hamiltonian. 
%  In practice, we must also
%  truncate 
%  %the number of 
%  %harmonics by 
%  $\lvert m\rvert\leq m_{\rm max}$,
%  where  $m$ counts the number of the harmonic process (exchange of energy packets of $\omega_{M}$).
%   With the onset of the time-dependence the time-independent Hamiltonian will be renormalized and the new static part.

Note that $\mathbf{H}_m$ separates into a 
{\it time-independent} part for $m=0$ and $m \neq 0$ (for finite $R_0$, including 
a shift due to $R_{M}\neq0$), 
%as well as
and a  time-dependent interaction.
Thus, while 
%It is interesting to point out that due to the nonzero presence of 
$R_{M}$ is  related to the %emergence of 
 time-dependent interaction, there is a static contribution 
from that term as well. 
% While Hamiltonian ${H}(t)$
% accounts for the static dressing
% via matter-matter (two-atom) interactions, 
%  Eq.~\eqref{H_m} dressed the entire
% two-atom system with periodic mechanical
% oscillations.
%  Similar expressions are widely considered for single quantum systems, including field-driven TLSs 
%  around the region of avoided level 
%  crossings~\cite{Ashhab_Two-level_2007}.
% For numerical calculations,
% the time-independent 
% %Hamiltonian, 
% ${H}_0$, is first diagonalized with the AC-shifted  
% dressed eigenbasis $\{E_j,\vert j\rangle\}$ for $j\in\{G^{\rm renorm},L^{\rm renorm},U^{\rm renorm},T^{\rm renorm}\}$
% obtained from:
% ${H}_{0}\vert j\rangle=E_j\vert j\rangle$ which satisfies the conditions $\langle j\vert j'\rangle=\delta_{jj'}$ and $\sum_{jj'}\vert j\rangle\langle j'\vert={\bf 1}$; 
% here
%  $E_j$ ($\lvert j\rangle$) are 
% shifted eigenenergies (eigenstates) renormalized [substituting $g\to g_0$ in E.~\eqref{EigEng}] by the presence of the nonzero $R_{M}$,
% since  the time-independent portion of the 
% %system 
% Hamiltonian is ${H}_0$ from Eq.~\eqref{H_m},
% and not ${H}\equiv{H}(t=0)$ in Eq.~\eqref{H}. 
% This 
% also 
% %helps to 
% ensures that we use the
% correct static states of the joint two-atom system
% in the presence of driving.
Solving the time-dependent equation of motion (which is similar to the time-dependent Schr\"{o}dinger equation with the Floquet theory), 
% $\ii\partial_t\vert\psi(t)\rangle={H}(t)\vert\psi(t)\rangle$, 
yields 
$\boldsymbol{B}_\alpha(t)=\mathrm{e}^{-\ii\varepsilon_\alpha t}\boldsymbol{F}_\alpha(t)$,
where $\varepsilon_\alpha$ 
is
the Floquet 
quasienergy~\cite{Nikishov_Quantum_1964}, and the Floquet mode $\boldsymbol{F}_\alpha(t)$ is  
$T$-periodic~\cite{Floquet_FloquetTheory_1883,Chicone_Ordinary_2006}.
 The 
 Floquet solutions, $\{\boldsymbol{B}_\alpha(t)\}$, form a complete basis for any value of $t$, 
 thus $\boldsymbol{B}(t)=\sum_\alpha c_\alpha \boldsymbol{B}_\alpha(t)$, where
 $c_\alpha=\boldsymbol{F}_\alpha^T\boldsymbol{B}_\alpha(0)$, with $\boldsymbol{F}_\alpha\equiv\boldsymbol{F}_\alpha(0)$.
Resonances 
occur at differences between Floquet quasienergies~\cite{Shirley_Solution_1965}.

To compute the Floquet modes~\cite{Akbari_Floquet_2025}, we use a Fourier series expansion of
$\boldsymbol{F}_\alpha(t)=\sum_{l\in\mathbb{Z}}\mathrm{e}^{\ii l\omega_{\rm M}t}\boldsymbol{F}_{\alpha l}$, where the Fourier coefficient vectors $\boldsymbol{F}_{\alpha l}$ are 
{\it Floquet sidebands}.
Although the dynamical matrix is $4\times4$ so that one might expect nominally $4$ eigenenergies and thus 4 quasienergies confined within a $\omega_M$ energy range, we only see 2 lines corresponding to the two modes (the determinant has only two real roots, the polariton resonances) as the Floquet theory is applied to the equation of motion (not the Hamiltonian) representing the effective transitions--from the dressed polariton states to the dressed ground state, within the system~\cite{Akbari_Floquet_2025}. Indeed, the coupling between these quasienergy lines is drive-dependent and for small enough $\eta$, $R_M$ and $\omega_M$, the drive might not have enough push to couple all the manifolds of the energy states [see the right-most part of Fig.~\ref{fig: schematics}(b)], and thus, only a limited number of them come into a Brillouin zone to make a coupling and transition possible.
This will also reveal the dynamics of the two dressed polariton modes, that is equivalent in classical and quantum pictures~\cite{Hughes_Reconciling_2024}.

Although we have made several key approximation in the theory above, we will see later that the resonances are in very good agreement with the full electrodynamic simulations.
To be specific, 
PyCharge employs a full wave approach without any approximation based on nonperturbative, linear electromagnetic theory (derived from Maxwell's equations), so it can capture the nondipolar, and non-Markovian
aspects of the coupling, as well as the damping.
However, the Floquet analysis above properly accounts for the main effect of the modulated dipole-dipole interaction including the intrinsic asymmetry stepping
from the $1/R^3(t)$ term, while giving a solid theoretical background to the main effects that we expect to see when we modulate the dipole-dipole coupling with a periodic drive. Moreover, the analysis can also be used to look at 
quantum effects for coupled two-level systems, which yields
different spectral signatures than the coupled harmonic oscillator problem (primarily due to saturation effects and biexcitonic coupling).
This will be the subject of future work. 

% , also having the capability of including the nonflat bath effects to some degree by the phenomenological damping rate of the Lorentzian model if $\gamma(t)$ is satisfactorily introduced. On the other hand, the Floquet analysis is based on the time-dependent solution of a classical/quantum (equation of motion)...TBC   

%\clearpage\newpage
\section{Results and Simulations}\label{sec: results}

\begin{figure}[th]
    \centering
    \includegraphics[width=0.84\linewidth]{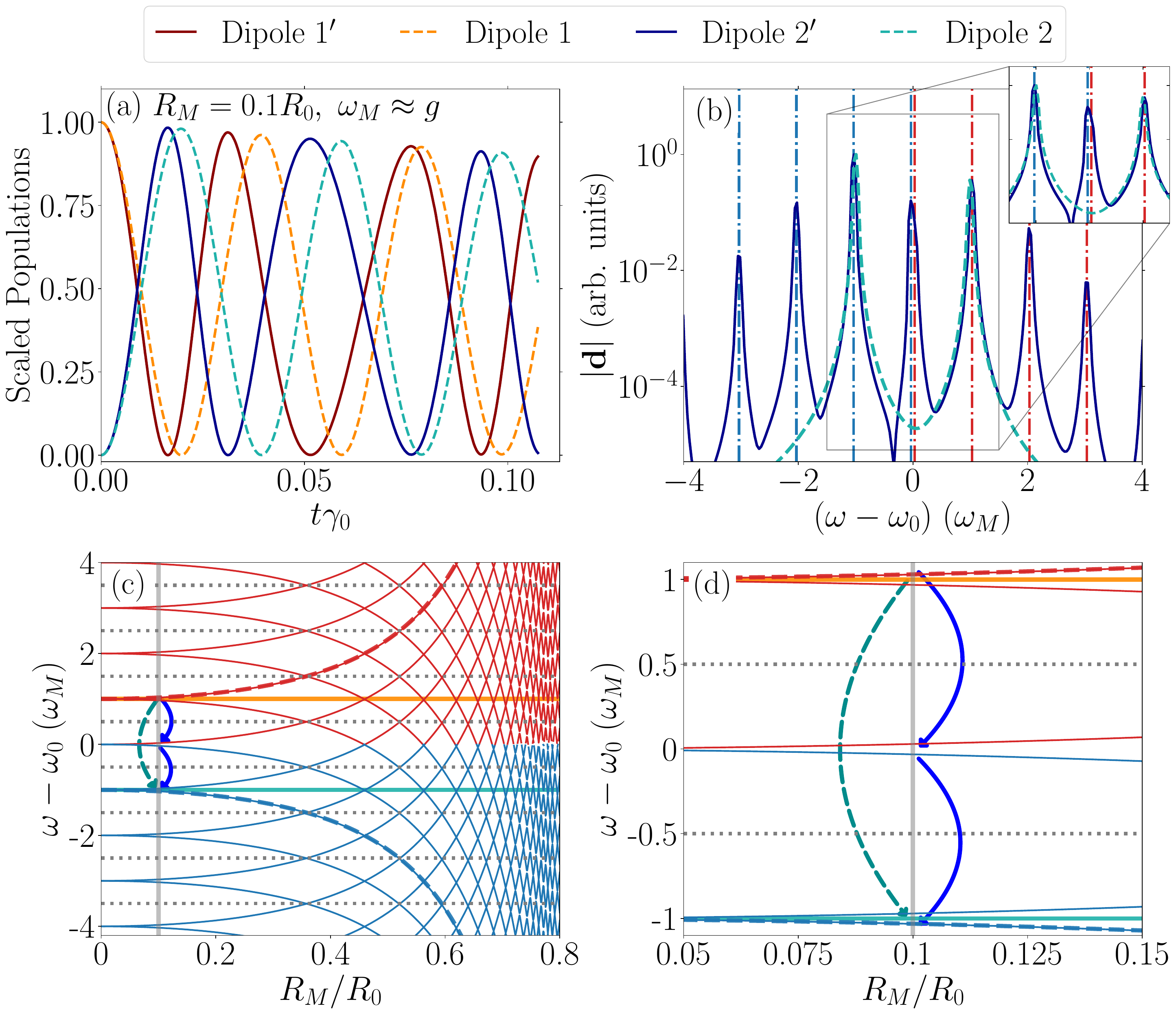}
     \caption{\textbf{PyCharge simulations of mechanically dressed dipole-dipole interactions and corresponding Floquet eigenfrequencies.}
    Two-coupled LOs ($\eta=g/\omega_0 = 0.00136$) with (Dipole 1$^\prime$: solid dark-red and 2$^\prime$: solid dark-blue) and without (Dipole 1: dashed orange and 2: dashed light-blue) mechanical motion, simulated by Eq.~\eqref{eq: mechanical motion}. (a) The normalized (scaled) populations for the dipoles are plotted under driving amplitude $R_M={0.1}R_0$ and frequency $\omega_M = g$, where Dipole 1 and 1$^\prime$ are initially excited, and Dipole 2 and 2$^\prime$ are not. These are obtained from Eq.~\eqref{eq: scaled population} using the total energy calculated by Eq.~\eqref{eq: total energy}, for the dipole moments in the time domain. Panel (b) gives the corresponding dipole frequency-domain spectra (Fourier transform of the time-dependent dipole moments) of the second dipole when its COM is stationary (dashed light-blue) or dynamic (solid dark-blue), where the transparent vertical blue
    and red dash-dotted lines are the Floquet quasienergies associated with their corresponding values [cf. panel (c), and also Fig.~\ref{fig: schematics}(b)] . In panel (c), the variation of the Floquet quasienergies versus the normalized amplitude of dynamical relative distance is plotted, where we also  see the dc eigenenergies (lower polariton: transparent thick solid light-blue line, upper polariton: transparent thick solid orange line) as well as the renormalized-dc eigenenergies (renormalized lower polariton: transparent thick solid blue line, renormalized upper polariton: transparent thick solid red line). Panel (d) is an inset of panel (c) that depicts a magnified zone of the energy scale near the parameter space used for the PyCharge results. The inset of panel (b) shows a magnification of the major (first-order perturbation) peaks, corresponding to the transitions shown by the arrows with the same line styles (i.e., dashed light-blue for the stationary Dipole 2 and solid dark-blue lines for the dynamic Dipole 2') in panels (c) and (d), in which the vertical transparent gray line highlights the desired parameter values; the higher-order peaks are repeated through all the Brillouin zones spaced by the integer of the drive frequency shown by the horizontal gray dotted lines. The free-space energy decay rate $\gamma_0$ of the dipole is 21.48 GHz ($R_0=50\,$nm, $y_0/R_0 = 0.02$, $q=10e$, initial dipole moment: $d_0=qy_0$, $\omega_0= 200\,$THz). 
}
    \label{fig: PyCharge simulation 1}
\end{figure}

\begin{figure}[b]
    \centering
    \includegraphics[width=0.84\linewidth]{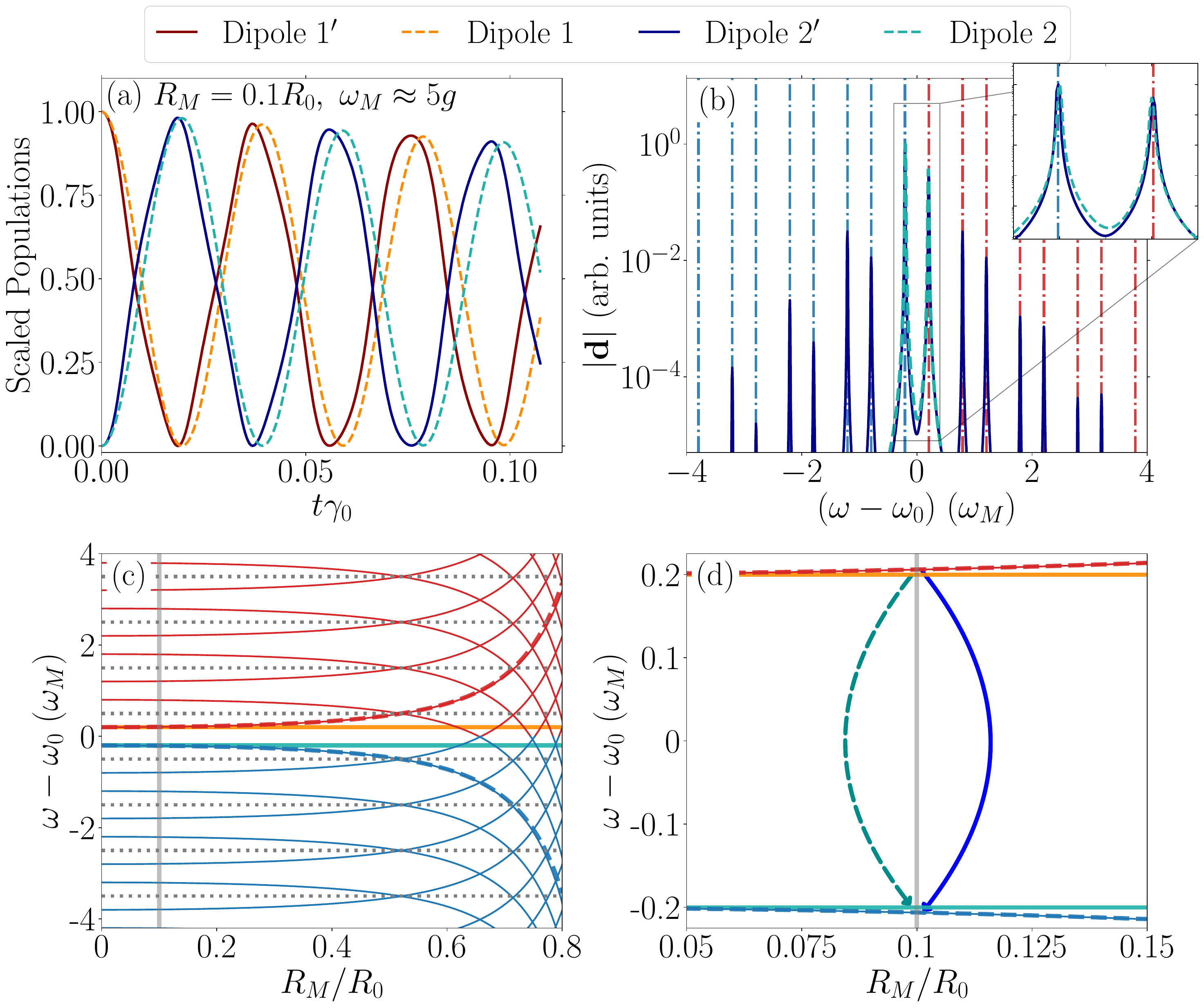}
    \caption{ Same as Fig.~\ref{fig: PyCharge simulation 1}, but with $R_M={0.1}R_0$ and $\omega_M = 5g$. 
    % The inset shows a magnification of the major (first-order perturbation) peaks.
    }
    \label{fig: PyCharge simulation 2}
\end{figure}

\begin{figure}[ht]
    \centering
    \includegraphics[width=0.84\linewidth]{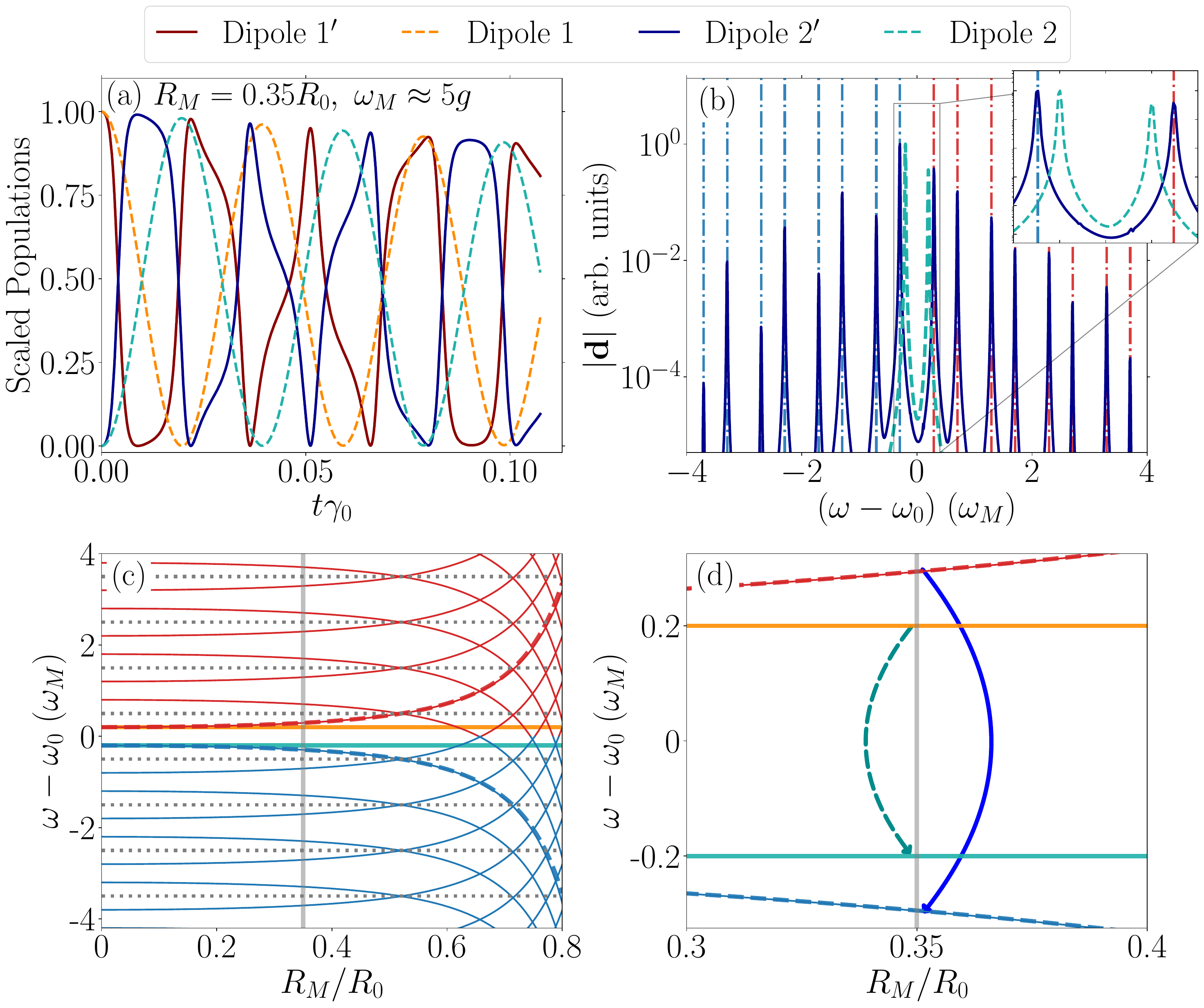}
    \caption{Same as Fig.~\ref{fig: PyCharge simulation 2}
    (again with $\omega_M = 5g$), but with $R_M={0.35}R_0$, where the vertical transparent grey line in panels (c) and (d) highlights this new parameter space. 
    % The inset shows a zoom in of the major (first-order perturbation) peaks.
    }
    \label{fig: PyCharge simulation 3}
\end{figure}

In this section, we show the results of the PyCharge simulations in time and frequency domains, and show its agreement with the Floquet theory. We let the first dipole be initially excited (internal charges are within their maximum relative distance) and the second one initially in the ground state (internal charges are within their minimum relative distance).
In Figs.~\ref{fig: PyCharge simulation 1}, \ref{fig: PyCharge simulation 2} and \ref{fig: PyCharge simulation 3}, we first investigate the time evolution of the {\it scaled} population of the excited states of the two coupled LOs ($s$-dipoles) in their panels (a), then show the Fourier transformation of the time-dependent dipole moments that shows the spectral peaks of the transitions in their panels (b), and finally show their corresponding Floquet quasienergy diagrams in panels (c-d) to explain that the transitions occur in the differences of the quasienergies. We use a fixed value of the (static) coupling between the dipoles of $g/\omega_0 = 0.00136 = g$, which is large enough to easily resolve the dipole-dipole splitting and coherent oscillations, since $g\gg \gamma_0$.
In addition,  we 
also choose 
  various values for the mechanical drive's frequency and dynamical coupling. Specifically,  we first use $R_M=0.1R_0$ and a rather low drive frequency of $\omega_M=g$, which is already large enough to  
  show clear new resonances, as shown in
 % \ymc{have we explain why do we choose these driving frequency in the beginning? Since the driving frequency we chose that are relative large compared with~\cite{Ferreri_Spontaneous_2019} and~\cite{Glaetzle_Single_2010} and also limit $\tilde{\alpha}\leq 0.8$} 
 Fig.~\ref{fig: PyCharge simulation 1}; 
 in addition, we investigate $R_M=0.1R_0$ and a higher drive frequency of $\omega_M=5g$ in Fig.~\ref{fig: PyCharge simulation 2}, and we increase the dynamical coupling by $R_M=0.35R_0$ and a rather high drive frequency of $\omega_M=5g$ in Fig.~\ref{fig: PyCharge simulation 3}.
Below we discuss each of these three regimes.

Figure~\ref{fig: PyCharge simulation 1}(a) shows the PyCharge simulation result of the (scaled) population of the excited states of the two coupled dipoles with (Dipole 1$^\prime$ and $2^\prime$, solid lines) and without (Dipole 1 and 2, dashed lines) mechanical oscillations, plotted as a function of time with the following mechanical parameters:
 $R_M = 0.1R_0$ and $\omega_M\approx g$ ($T=0.079\gamma_0^{-1}$),
where $T$ refers a cycle of mechanical oscillation period.
The time is scaled by the free-space decay rate ($\gamma_0$ = 21.48 GHz). Dipole 1 and 1$^\prime$ [$y_1(0)=1\,$nm] are initially excited, while Dipole 2 and 2$^\prime$ are not [$y_2(0)\approx 0$], corresponding to the schematic diagram in Fig.~\ref{fig: schematics}(a). All of the LOs have a natural angular frequency $\omega_0$ of 200 THz. To ensure very good accuracy, the numerical simulation runs over 10 million time iterations (125,663 periods of $\omega_0$) with a time step $\Delta t=10^{-17}\,$s [$(\omega_0\Delta t)^{-1}\approx80$]. The dipole-dipole initial separation is $R_0=50\,$nm ($y_0/R_0=0.02$) along the $x$-axis, with charge magnitude $q=10e$, effective mass $m_{\rm eff}=\hbar/2\omega_0 y_0^2$ and static coupling strength $g\equiv g_{12}=g_{21} = 79.7\gamma_0$ ($\tau_{dd} = \pi g^{-1}=0.039\gamma_0^{-1}$, where $\tau_{dd}$ represents the period of a cycle in which a photon is emitted and subsequently re-absorbed by the same dipole, or simply the round trip time in a classical picture). 
The time domain simulations show an oscillatory decay (caused by dipole-dipole interactions) of the populations to a steady state, where, with the onset of the mechanical oscillation, the phase of the oscillatory decay changes dynamically; thus, the time interval between the same peaks of the moving and stationary dipoles is sometimes shorter and sometimes longer. However, this oscillatory decay behavior for the mechanically oscillating dipoles is always faster than the stationary dipoles; this indicates the emergence of higher frequency peaks in the spectral analysis [cf. spectral peaks in panel (b) and also panel (c)].

Figure~\ref{fig: PyCharge simulation 1}(b) displays the spectral peaks of the dipoles in the frequency domain (shown in a logarithmic scale) for the corresponding cases in Figs~\ref{fig: PyCharge simulation 1}(a), centered at the dipoles' transition frequency $\omega_0$ and normalized by $\omega_M$. The spectra are obtained by a fast Fourier transform (FFT) of the dipole moment $d_n(t)$ with a 
Hamming window function applied to enhance peak clarity and smoothness. We see that all of the resonance peaks from the field spectrum plots are separated by an integer number of the driving frequency $\pm\omega_M$ ($n$ integer) when $\omega_M\leq g$, as shown in Fig.~\ref{fig: PyCharge simulation 1}(c). The blue and red dot-dashed vertical lines in panel (b) are the same as the blue and red thin solid lines representing the Floquet quasienergies, calculated from Eq.~\eqref{H_m}, in panel (c), for the desired values of the parameter space pointed out by the vertical transparent gray line, and with a magnified view in panel (d). 
This spectral behavior contrasts significantly with that of two coupled LOs with a fixed COM, which produce only  two frequency peaks (the well known superradiant and subradiant states, separated
by $2g$, which also connect to Dicke states~\cite{Dicke_Coherence_1954}). As anticipated, mechanical oscillation can modify these states to a manifold of dressed (Floquet) states, shown in Fig.~\ref{fig: schematics}, where energy levels are separated by integer multiples of $\omega_M$.

Figure~\ref{fig: PyCharge simulation 1}(c) shows the Floquet quasienergies near $\omega_0$, scaled with $\omega_M$, where 
$\omega_M/\omega_0=\eta=0.00136$, for two coupled HOs as a function of normalized oscillating amplitude ($R_M/R_0$) for the same regime as that of panels (a) and (b).
Here, we include a truncated number of Brillouin zones (separated by the grey dotted lines). 
Construction of the Brillouin zones, which is dependent on the strength of the external drive, assists the coupling of the energy states by the external quanta; thus, with a limited (not enough) number of Brillouin zones, due to the small choices of drive frequency and amplitude, the complete intermixing of all the system's eigenenergy states might not be possible within a Brillouin zone. Moreover, within the small range of the drive amplitude, we do not yet observe the creation of effective anticrossings in the dynamical picture between the quasienergy levels, which would only occur  for sufficiently larger values of $R_M/R_0$.

In coupling scenarios where $\omega_M \leq g$, the spectral lines are initially separated by the integers of the driving frequency,
as demonstrated in  Fig.~\ref{fig: PyCharge simulation 1}(c), where we normally expect a single peak structure within the range of one $\omega_M$. We also see that at $\omega_M\sim g$ (and, in general, increasing the coupling strength via $g$ and $R_M$) we are well at the verge of entering into the regime of splitting the peak, as seen in panel (b), where also the initial splitting of the Floquet quasienergy lines almost begin to occur near the vertical grey line in panel (c); which is clearer in the magnified region in panel (d).
However, a split in the spectral lines with an internal separation of $2\eta$ is observed when $\omega_M> g$ (and, more effective for $\omega_M\gg g$), because the splitting of the quasienergy lines is now more effective and also they may undergo through an ac Stark shift. Additionally, the center of splits is still separated by an integer multiple of $\omega_M$ because still we encounter transition only between the self repetition  manifolds (going back and forth in different Brillouin zones from a state) of the polariton states only.

Figure~\ref{fig: PyCharge simulation 2} is similar to Fig.~\ref{fig: PyCharge simulation 1}, but with a larger driving frequency, $\omega_M\approx 5g$ ($T=0.016\gamma_0^{-1}$).
Comparing the results with (primed dipoles) and without mechanical oscillation (unprimed dipoles), i.e., comparing panel (a)'s in Figs.~\ref{fig: PyCharge simulation 1} and \ref{fig: PyCharge simulation 2}, the driving frequency clearly alters the phase of the population oscillations
in a different way due to time-dependent fluctuation in the dipole-dipole coupling strength. 
As shown in Fig.~\ref{fig: PyCharge simulation 2}(a), we observe an overall faster exchange of energy between Dipole 1$^\prime$ and 2$^\prime$ (mechanical) compared with Dipole 1 and 2 (non-mechanical), when the driving frequency is greater than the dipole-dipole frequency shift, $\omega_M=5g>g$, indicating an increase of the frequency shift.
In other words, for $\omega_M=g$ in Fig.~\ref{fig: PyCharge simulation 1}(a), we see that the populations of mechanical dipoles (1' and 2') oscillate sometimes faster, sometimes slower than the non-mechanical cases (1 and 2). However, once we have a larger driving frequency, the population exchange of mechanically moving dipoles are always faster than the non-mechanical case. This can imply that for $\omega_M\gg g$, one can expect an overall faster exchange of energy between Dipole 1$^\prime$ and 2$^\prime$ (mechanical).

Furthermore, with $\omega_M=5g>g$, we observe that the resonance peaks tend to exhibit a splitting into superradiant and subradiant states with internal separation around 2$g$, as illustrated in Fig.~\ref{fig: PyCharge simulation 2}(b). 
The inset of panel (b) provides an enlarged view of the peaks associated with the dressed polariton states. 
Additionally, the center of the splitting peaks is still separated by an integer of driving frequency. The vertical blue and red dot-dashed lines show these phenomena are aligned with our Floquet analysis, showing the
excellent agreement between Floquet analysis and PyCharge simulations.

We next increase the oscillating amplitude that can bring the dipoles much closer to each other to explore the role of the oscillating strength (of the mechanical drive), again with $\omega_M=5g$, as shown in Fig.~\ref{fig: PyCharge simulation 3},
where we now use a larger driving amplitude, $R_M= 0.35R_0$.
The change in the phase of population oscillations has not been significant, but the frequency of the population oscillations increases and decreases dramatically, as the effective $\tau_{dd}$ is significantly reduced when the dipoles are brought closer together, and 
significantly increases when the dipoles are farther apart. This causes the resonances to be modified dynamically, as shown in Fig.~\ref{fig: PyCharge simulation 3}(b-d).

We also find that the internal separation between resonance peaks got further split when $R_M/R_0$ is large enough. Our study indicates that the oscillating amplitude $R_M$ is able to influence the positioning of resonant frequencies, causing a splitting of the original peaks, as well as a change in the separation between the split peaks (when the amplitude is sufficiently large). 
The ensuing dynamic control is also corroborated by Floquet analysis, see panels (c) and (d) of Fig.~\ref{fig: PyCharge simulation 3}, where we see the renormalization of the dc eigenenergies (dashed lines) and the ac Stark shift of quasienergies (thin solid lines) are becoming largely impactful and are well distinguishable from the dc eigenenergies (thick solid lines). 

In Figs.~\ref{fig: PyCharge simulation 1}, \ref{fig: PyCharge simulation 2} and \ref{fig: PyCharge simulation 3}, Floquet quasienergies and states are obtained by analyzing the system's Floquet dynamical matrix corresponding to the equation of motion, which was constructed in an extended Hilbert space incorporating time periodicity, similar to the approach in Ref.~\cite{Akbari_Floquet_2025}, and their correspondence to spectral features  of the dipole moment $|{d}_n(\omega)|$. Hence, this approach confirms the presence of well-defined Floquet states, with spectral peaks appearing at integer multiples of the driving frequency. Additionally, the theoretical Floquet model was validated by the numerical simulations, providing insights into the effects of mechanical motion and dipole-dipole interactions on the system's energy structure. 

Additionally, a closer examination of the results in the Floquet analysis reveals that, when the parameter $R_M/R_0$ reaches sufficiently high values, the quasienergy states exhibit crossing/anticrossing behavior. The region of crossing and avoided crossing occurs more rapidly with increasing oscillating amplitude, resulting in a dispersion and merging of the resonances. This phenomenon is better illustrated in Fig.~\ref{fig: PyCharge simulation 3}(c). 
The region of crossing/anticrossing shifts toward  larger values of normalized driving amplitude as the frequency of the drive increases. Thus, the amplitude modulation may decreases as driving frequency increases, similar to the findings in~\cite{Glaetzle_Single_2010}, and the amplitude of this modulation decreases when the oscillation frequency of the mirror increases.

Finally, we remark that
we have also investigated the influence of the mechanical oscillation phase, along with the consequences of altering the signs of sine and cosine functions. We found that these effects are minimal when the mechanical frequency is sufficiently high. However, they can become more pronounced at lower driving frequencies,
which was not the domain of interest for this study.

\section{Conclusions}\label{sec: conclusion}

We have presented a direct numerical simulation of
moving point charge potentials,
using PyCharge, 
and studied how the (scaled) population dynamics  and dipole spectra
of two two-dipole coupled  LOs 
can be significantly influenced by periodic mechanical oscillations, in a non-trivial way. We also investigated how the spectral characteristics and resonance peaks 
vary with the mechanical perturbation parameters, $R_M/R_0$ and $\omega_M/g$, and 
showed how the resonance peaks positioning, spectral splitting of the peaks, as well as the separation between the split peaks change with an increased oscillating amplitude $R_M$. 

These results were fully corroborated with an intuitive and rigorous Floquet theory, where we added a periodic oscillation to the usual dipole-dipole coupling term (which can be derived from classical or quantum Maxwell equations), which couples two HOs through a 
time-dependent coherent coupling term.
Both the direct numerical and Floquet analysis demonstrated a reciprocal relationship between $R_M/R_0$ and $g$, with resonances showing a separation by an integer of driving frequency, namely $\pm l\omega_M$, and these resonances further split into superradiant and subradiant states when drive frequency becomes larger than the static dipole-dipole coupling. Significantly, in scenarios with sufficiently large values of both $R_M$ and $\omega_M$, there was a reduction in the radiative decay rate of the system, implying that mechanical effects can effectively enhance the coupling strength %constantly
and extend the lifetime of the coupling system. 

Related effects have been predicted in studies of an atom in front of an oscillating mirror (phonon-photon coupling)~\cite{Glaetzle_Single_2010,Ferreri_Spontaneous_2019}, and time-modulated photonic crystals (time-modulated band-gap)~\cite{Calajo_Control_2017}, where the sidebands in the emitted spectrum are separated from the atomic transition frequency by the modulation frequency of the plate or photonic lattice, $\omega_M$; however, the lateral peaks here were strongly asymmetric due to the different density of states at the edges of the photonic band gap~\cite{Calajo_Control_2017}. In contrast, for our case, which deals with finite-size dipoles in free space, the lateral peaks are symmetric because the photonic density of states (DOS) is essentially the same at frequencies of the peaks. 

The advantage of PyCharge is that we do not need to make 
any of the usual approximations (dipole approximation and rotating wave approximation) and although everything is classical in nature, we are still able to show clear analogies 
with quantum optical systems (quantum emitters and quantized fields)~\cite{Glaetzle_Single_2010,Ferreri_Spontaneous_2019}.
Additionally, we have demonstrated with a stronger oscillating strength or faster driving frequency, the resonance peaks can be further modified. We have gone beyond usual perturbative optomechanical studies, where they have linearized Taylor series expansion~\cite{Agarwal_Quantum_2012,Chiangga_Perturbation_2024,Primo_Quasinormal-mode_2020}.

Overall, our findings provide a deeper understanding of the interplay between light and matter with mechanical motion, 
using coupled atoms described by 
LOs and the self-consistent fields. This work lays the foundation for exploring more complex optomechanical systems, potentially leading to novel applications in
fundamental nano-optics,
quantum information processing, and sensing and communication protocols
(e.g., using coupled atoms as nano antennas).
Moreover, it showcases the opportunities
of using numerical solutions of retarded potentials and moving point charges and dipoles (using and extending the open-source PyCharge software), in contrast to the usual numerical Maxwell solvers with the physical fields
(such as through FDTD and finite-element solvers).

Finally, we remark
these classical studies can also be used to help develop more advanced quantum models, since a classical-quantum correspondence is only expected in some limit, e.g., linear response with no saturation effects (HO models). Thus, future work, for example, could look at the dynamical coupling between
dipoles treated as
two-level systems, where the eigenstates 
also contain a two quanta excitation, and saturation and nonlinear effects can become important. 

\acknowledgements
 This work was supported by the Natural Sciences and Engineering Research Council of Canada (NSERC),
 the National Research Council of Canada (NRC),
 the Canadian Foundation for Innovation (CFI), and Queen's University, Canada.
%%%%%%%%%%%%%%%%%%%%%%%%%%%%%%%%%%%%%%%%%%%%%%%%%%%%%%%%
% section: bibliography
%%%%%%%%%%%%%%%%%%%%%%%%%%%%%%%%%%%%%%%%%%%%%%%%%%%%%%%%
% \bibliography{thesis.bib}
\bibliography{Refs}

\clearpage
\newpage
% %%%%%%%%%%%%%%%%%%%%%%%%%%%%%%%%%%%%%%%%%%%%%%%%%%%%%%%%
% % section: appendices
% %%%%%%%%%%%%%%%%%%%%%%%%%%%%%%%%%%%%%%%%%%%%%%%%%%%%%%%%
\appendix
\section{Electrodynamics of interacting and oscillating dipoles in PyCharge}
\label{sec:dipole-fields}

In classical electrodynamics, the 
analytical solution for the 
electric and magnetic fields at position ${\bf R}$ from the COM of an idealized electric dipole at time $t$,
in a homogeneous medium, is given by~\cite{Jackson_Classical_1999,Novotny_Principles_2012} 
\begin{align}\label{eq: classical E field}
\mathbf{E}(\mathbf{R},t) = \frac{1}{4\pi\epsilon_0 \epsilon_{\rm B}}\Bigg[k^2(\mathbf{\hat{R}}\times{\mathbf{d}})\times\mathbf{\hat{R}}\frac{e^{ikR}}{R}+
%\\
\left[3(\mathbf{\hat{R}}\cdot{\mathbf{d}})\mathbf{\hat{R}}-{\mathbf{d}}\right]\Bigg(\frac{1}{R^3}-\frac{ik}{R^2}\Bigg)e^{ikR}\Bigg],
\end{align}
and 
\begin{equation}\label{eq: classical B field}
\mathbf{B}(\mathbf{R},t) = \frac{\mu_0}{4\pi}\left[ck^2(\mathbf{\hat{R}}\times\mathbf{d})\left(1-\frac{1}{ikR}\right)\right]\frac{e^{ikR}}{R},
\end{equation}
where $k = \omega/c$, $\mathbf{d}=\mathbf{d}_0e^{-i\omega t}$ is the time-dependent harmonic oscillator
dipole moment, $R=|\mathbf{R}|$, and $\mathbf{\hat{R}}=\mathbf{R}/R$ is unit vector of $\mathbf{R}$.
Note here that the two-body charges constituting the dipole oscillate but the COM of the dipole is at rest.
In free space, then
$\epsilon_{\rm B}=1$.

When the COM of a dipole is in motion, one still can use Eqs.~\eqref{eq: classical E field} and \eqref{eq: classical B field} to calculate the electromagnetic fields from a moving charge density by changing ${\bf R}\to\mathbf{R}(t)=\mathbf{v}t$ provided that the motion is with a constant velocity $\mathbf{v}$ throughout time (nonaccelerating).
However, when the motion is with time-varying velocity $\mathbf{v}(t)$, the contributions from the retarded and advanced time difference between the source and the observation point play a role and hence contribute to radiation. The EM field of an accelerated charge density can be calculated via well-known Li\'{e}nard-Wiechert potentials~\cite{Jackson_Classical_1999}. 

% \subsection{Li\'{e}nard-Wiechert potentials for moving dipoles}
 
The charge and current densities of a point charge $q$ at the position $\mathbf{r}_s(t)$ with velocity $\mathbf{\dot{r}}_s(t)$ are, respectively,
\begin{equation}\label{eq: charge_d}
    \rho(\mathbf{r},t) = q\delta[\mathbf{r}-\mathbf{r}_s(t)],
\end{equation}
and
\begin{equation}\label{eq: current_d}
    \mathbf{j}(\mathbf{r},t) = q\dot{\mathbf{r}}_s(t)\delta[\mathbf{r}-\mathbf{r}_s(t)].
\end{equation}

The scalar and vector potentials associated with a moving point charge in the Lorenz gauge, known as the Li$\rm\acute{e}$nard-Wiechert potentials~\cite{Wiechert_Elektrodynamische_1901,Griffiths_Introduction_2013}, are derived from Maxwell's equations,
 and can be written as
\begin{align}\label{eq: scalar potential}
    \mathbf{\Phi}(\mathbf{r},t) = \frac{1}{4\pi\epsilon_0}\int \mathrm{d}t_r\,\frac{\rho(\mathbf{r},t_r)}{R_s}=\frac{q}{4\pi\epsilon_0}\left[\frac{1}{{\kappa}(t_r) {R}_s}\right],
\end{align}
and 
\begin{align}\label{eq: vector potential}
    \mathbf{A}(\mathbf{r},t) =\frac{\mu_0}{4\pi}\int \mathrm{d}t_r\,\frac{\mathbf{j}(\mathbf{r},t_r)}{R_s}= \frac{\mu_0 q}{4\pi}\left[\frac{\mathbf{\dot{r}}_s(t_r)}{{\kappa}(t_r){R}_s}\right] 
    %\\
    = \frac{\mathbf{\dot{r}}_s(t_r)}{c^2}\mathbf{\Phi}(\mathbf{r},t),
\end{align}
where $R_s=|\mathbf{r}-\mathbf{r}_s(t_r)|$ is the distance from source points, and ${\kappa}(t_r)=1-\mathbf{n}_s(t_r)\cdot\boldsymbol{\beta}_s(t_r)$, such that $\mathbf{n}_s=\mathbf{R}_s/R_s$ is the unit vector from the position of the point charge to the field point and $\boldsymbol{\beta}_s(t_r) = \mathbf{\dot{r}}_s(t_r)/c$ is the velocity of the point charge expressed as a fraction of the speed of light. The quantity in brackets, evaluated at the retarded time $t_r$, is given by
\begin{equation}\label{eq: retarded time}
    t_r = t-\frac{R_s(t_r)}{c},
\end{equation}
where the delay time can be %numerically 
solved numerically, e.g., by the secant method or Newton's method.

We can subsequently calculate the electric and magnetic fields of a point charge in arbitrary motion, using  the solution of the Li$\rm\acute{e}$nard-Wiechert potentials. 
The calculation of electromagnetic fields and potentials of dipoles in PyCharge applies the principle of superposition of classical electrodynamics, which is the sum of individual contributions of each source. Each contribution is calculated along discretized spatial grids at a specific time. The physical (gauge-invariant) relativistically-correct, time-varying total electric field is given by a sum of the electric Coulomb (velocity) and radiation (acceleration) fields \cite{Griffiths_Introduction_2013,Jackson_Classical_1999},
${\bf E}(\mathbf{r},t)={\bf E}_{\rm Coul}(\mathbf{r},t)+{\bf E}_{\rm rad}(\mathbf{r},t) = -\boldsymbol{\nabla}\mathbf{\Phi}-\frac{\partial \mathbf{A}}{\partial t}$, where
\begin{align}\label{eq: E_coul}
    \mathbf{E}^{\rm Coul}(\mathbf{r},t) = \frac{q}{4\pi\epsilon_0}\frac{R_s(t_r)}{(\mathbf{R}_s(t_r)\cdot\mathbf{u}(t_r))^3}
    %\\
    \left[\left(c^2-|\mathbf{v}_s(t_r)|^2\right)\mathbf{u}(t_r)\right],
\end{align}
and 
\begin{align}\label{eq: E_rad}
    \mathbf{E}^{\rm rad}(\mathbf{r},t) =\frac{q}{4\pi\epsilon_0}\frac{R_s(t_r)}{(\mathbf{R}_s(t_r)\cdot\mathbf{u}(t_r))^3}
    %\\
    \left[\left(\mathbf{R}_s(t_r)\times(\mathbf{u}(t_r)\times\mathbf{a}(t_r)\right)\right],
\end{align}
with $\mathbf{v}_s(t_r)$ the velocity of the charge, $\mathbf{a}(t_r)$ the acceleration of the charge, and
\begin{equation}\label{eq: u}
    \mathbf{u}(t_r) = c\mathbf{n}_s(t_r)-\mathbf{v}_s(t_r),
\end{equation}
where $\mathbf{r}_{s}(t_r)$ is the position of the charge at the retarded time, and $\mathbf{r}$ is the current position of the charge.

\section{Dipole-coupled harmonic oscillator theory using two electric dipoles in the near field regime}
\label{sec:dip-dip}

\subsection{Quantum optical approach for two coupled dipoles}

For any general medium,
the Born-Markov 
reduced master equation 
for two coupled dipoles (treated as quantized HOs), in the rotating wave approximation, is
\cite{PhysRevA.66.063810,Angelatos_Entanglement_2015}
\begin{align}
    \frac{\mathrm{d}\rho}{\mathrm{d}t} = ~ -&\ii\sum_{n=1,2}\omega'_n\left[b^\dagger_{n}b_{n},\rho \right] 
    -\ii[g_{12} (b^\dagger_1 b_2 
    + b^\dagger_2 b_1),\rho] 
    + \sum_{n= 
    1,2}\sum_{m = 
    1,2}\frac{\gamma_{nm}}{2}\left[2b_n\rho b^\dagger_{m}-b^\dagger_n b_{m}\rho -\rho b^\dagger_n b_{m}\right]  ,
    \label{eq: two atom master with gain}
\end{align}
%\end{widetext}
where we define $\omega'_n=\omega_n+\delta_{n}$,
and assume 
 equal dipole resonances
and dipole moments, so that
$\omega_1' =\omega_2'=\omega_0$ (including a trivial Lamb shift, $\delta_{n}$). 
Thus all the coupling rates above are implicitly evaluated
at $\omega_0$. These rates are defined in the main text,
and explicitly, for the 
relevant coherent dipole-dipole coupling term,
we have 
\begin{equation} % \label{eq: frequency_shift}
g_{12} = -\frac{{\bf d}_{1} \cdot {\rm Re} {\bf G}({\bf R}_1,{\bf R}_{2},\omega_{0}) \cdot {\bf d}_{2}}{\epsilon_0 \hbar}.
\label{g_ij2}
\end{equation}

Neglecting the dissipation, then we obtain the relevant effective Hamiltonian,
\begin{equation}
H_{\rm eff}^{\rm QM}
= \hbar \omega_0 b_1^\dagger b_2 
+ \hbar \omega_0 b_2^\dagger b_2
+
\hbar g(b_1^\dagger b_2
+ b_2^\dagger b_1),
\end{equation}
where $g \equiv g_{12}$.
Furthermore, if not adopting a rotating wave approximation, then
\begin{equation}
H_{\rm eff}^{\rm QM}
= \hbar \omega_0 b_1^\dagger b_1 
+\hbar \omega_0 b_2^\dagger b_2
+\hbar g(b_1+b_1^\dagger)(b_2+b_2^\dagger).
\label{eq:dd3}
\end{equation}
Using this to obtain the 
 Heisenberg equation of motion, with $\mathrm{d}\boldsymbol{B}/\mathrm{d}t=-\ii\mathbf{H}\,\boldsymbol{B}$, where $\boldsymbol{B}=[b_1, b_2, b_1^\dagger, b_2^\dagger]^T$,
yields the main matrix equation in the main text, for which we use
to construct a Floquet solution. Moreover, in the case of 
dipoles in free space,
then 
\begin{equation}
\hbar g = \frac{d_1 d_2}{4\pi \epsilon_0 R^3}, 
\end{equation}
which is justified in more detail below, along
with a connection to
a classical model of dipole-dipole coupling, where a direct correspondence is shown.

\subsection{Direct classical approach from the 
Green's function formalism with two dipole oscillators}

The same result
(resonances and Hamiltonian matrix form) can be obtained directly from the classical electromagnetic theory and the Green's function formalism.
The interaction between two dipolar charge distribution are often described via the reciprocating electrostatic Coulomb force between them identified by, within the dipole approximation (and near field regime), but one can also use the Green's function formalism in the presence of scattering bodies to find the total electric field in either of the dipoles.

The exact dyadic Green's function  solution $\mathbf{G}^{(2)}(\mathbf{r},\mathbf{r}',\omega)$ with the source point at position $\mathbf{r}'$, two point-dipoles (scatterers) at positions
${\bf R}_1$ and ${\bf R}_2$, and observation point at position $\mathbf{r}$,
in free space, is obtained via the self-consistent solution
to the Dyson equation~\cite{Martin_Electromagnetic_1998}, including the scattering properties of all scatterers in the background medium, via
\begin{equation}
    \begin{split}
        \mathbf{G}^{(2)}(\mathbf{r},\mathbf{r}',\omega)&=\mathbf{G}^{\rm B}(\mathbf{r},\mathbf{r}',\omega)+\sum_{n=1,2}\mathbf{G}^{\rm B}(\mathbf{r},\mathbf{R}_n,\omega)\cdot\boldsymbol{\alpha}_n(\omega)\cdot \mathbf{G}^{(2)}(\mathbf{R}_n,\mathbf{r}',\omega),
    \end{split}
\end{equation}
where $\mathbf{G}^{\rm B}(\mathbf{r},\mathbf{r}',\omega)$ is the Green's tensor of the background medium and $\boldsymbol{\alpha}_n(\omega)$ is the polarizability tensor of the $n$th dipole.
We are interested in finding the total electric field at one of the dipoles from the other one, e.g. $\mathbf{G}^{(2)}(\mathbf{R}_1,\mathbf{R}_2,\omega)$, containing the scattering effect of both dipoles.
The superscript `(2)' indicates we have the solution with two dipoles includes. 
 
Here, we have adopted an exact iterative
approach to write a series of Dyson equations given by 
\cite{Kristensen_Decay_2011}, as
\begin{equation}
    \begin{split}
        \mathbf{G}^{(2)}(\mathbf{R}_1,\mathbf{R}_2,\omega)&=\mathbf{G}^{(1)}(\mathbf{R}_1,\mathbf{R}_2,\omega)+\mathbf{G}^{(1)}(\mathbf{R}_1,\mathbf{R}_2,\omega)\cdot\boldsymbol{\alpha}_2(\omega)\cdot \mathbf{G}^{(2)}(\mathbf{R}_2,\mathbf{R}_2,\omega),
    \end{split}
\end{equation}
where 
\begin{equation}
    \begin{split}
        &\mathbf{G}^{(1)}(\mathbf{R}_1,\mathbf{R}_2,\omega)=\mathbf{G}^{(0)}(\mathbf{R}_1,\mathbf{R}_2,\omega)+\mathbf{G}^{(0)}(\mathbf{R}_1,\mathbf{R}_1,\omega)\cdot\boldsymbol{\alpha}_1(\omega)\cdot \mathbf{G}^{(1)}(\mathbf{R}_1,\mathbf{R}_2,\omega)
        \\
        &\to\,\mathbf{G}^{(1)}(\mathbf{R}_1,\mathbf{R}_2,\omega)=\lsb\mathbf{I}-\mathbf{G}^{(0)}(\mathbf{R}_1,\mathbf{R}_1,\omega)\cdot\boldsymbol{\alpha}_1(\omega)\rsb^{-1}\cdot\mathbf{G}^{(0)}(\mathbf{R}_1,\mathbf{R}_2,\omega),
        % \equiv\widetilde{\mathbf{G}}_{\mathbf{R}_2,\mathbf{R}_1}(\omega),
    \end{split}
\end{equation}
and, 
\begin{equation}
    \begin{split}
        &\mathbf{G}^{(2)}(\mathbf{R}_2,\mathbf{R}_2,\omega)=\mathbf{G}^{(1)}(\mathbf{R}_2,\mathbf{R}_2,\omega)+\mathbf{G}^{(1)}(\mathbf{R}_2,\mathbf{R}_2,\omega)\cdot\boldsymbol{\alpha}_2(\omega)\cdot \mathbf{G}^{(2)}(\mathbf{R}_2,\mathbf{R}_2,\omega)
        \\
        &\to\, \mathbf{G}^{(2)}(\mathbf{R}_2,\mathbf{R}_2,\omega)=\lsb\mathbf{I}-\mathbf{G}^{(1)}(\mathbf{R}_2,\mathbf{R}_2,\omega)\cdot\boldsymbol{\alpha}_2(\omega)\rsb^{-1}\cdot\mathbf{G}^{(1)}(\mathbf{R}_2,\mathbf{R}_2,\omega),
    \end{split}
\end{equation}
with 
% \begin{equation}
%     \begin{split}
%         &\mathbf{G}^{(1)}(\mathbf{r}_1,\mathbf{r}_1,\omega)=\mathbf{G}^{(0)}(\mathbf{r}_1,\mathbf{r}_1,\omega)+\mathbf{G}^{(0)}(\mathbf{r}_1,\mathbf{r}_1,\omega)\cdot\boldsymbol{\alpha}_1(\omega)\cdot \mathbf{G}^{(1)}(\mathbf{r}_1,\mathbf{r}_1,\omega)
%         \\
%         &\to\, \mathbf{G}^{(1)}(\mathbf{r}_1,\mathbf{r}_1,\omega)=\lsb\mathbf{I}-\mathbf{G}^{(0)}(\mathbf{r}_1,\mathbf{r}_1,\omega)\cdot\boldsymbol{\alpha}_1(\omega)\rsb^{-1}\cdot\mathbf{G}^{(0)}(\mathbf{r}_1,\mathbf{r}_1,\omega)\equiv\widetilde{\mathbf{G}}_{\mathbf{r}_1,\mathbf{r}_1}(\omega),
%     \end{split}
% \end{equation} 
% and,
\begin{equation}
    \begin{split}
        \mathbf{G}^{(1)}(\mathbf{R}_2,\mathbf{R}_2,\omega)&=\mathbf{G}^{(0)}(\mathbf{R}_2,\mathbf{R}_2,\omega)+\mathbf{G}^{(0)}(\mathbf{R}_2,\mathbf{R}_1,\omega)\cdot\boldsymbol{\alpha}_1(\omega)\cdot \mathbf{G}^{(1)}(\mathbf{R}_1,\mathbf{R}_2,\omega)
        \\
        % &=\mathbf{G}^{(0)}(\mathbf{r}_2,\mathbf{r}_2,\omega)+\mathbf{G}^{(0)}(\mathbf{r}_2,\mathbf{r}_1,\omega)\cdot\boldsymbol{\alpha}_1(\omega)\cdot\lsb\mathbf{G}^{(0)}(\mathbf{r}_1,\mathbf{r}_2,\omega)+\mathbf{G}^{(0)}(\mathbf{r}_1,\mathbf{r}_1,\omega)\cdot\boldsymbol{\alpha}_1(\omega)\cdot \mathbf{G}^{(1)}(\mathbf{r}_1,\mathbf{r}_1,\omega)\rsb
        % \\
        % &=\mathbf{G}^{(0)}(\mathbf{r}_2,\mathbf{r}_2,\omega)+\mathbf{G}^{(0)}(\mathbf{r}_2,\mathbf{r}_1,\omega)\cdot\boldsymbol{\alpha}_1(\omega)\cdot\lcb\mathbf{G}^{(0)}(\mathbf{r}_1,\mathbf{r}_2,\omega)+\mathbf{G}^{(0)}(\mathbf{r}_1,\mathbf{r}_1,\omega)\cdot\boldsymbol{\alpha}_1(\omega)\right.
        % \\
        % &\hspace{7.5cm}\left.\cdot\lsb\mathbf{I}-\mathbf{G}^{(0)}(\mathbf{r}_1,\mathbf{r}_1,\omega)\cdot\boldsymbol{\alpha}_1(\omega)\rsb^{-1}\cdot\mathbf{G}^{(0)}(\mathbf{r}_1,\mathbf{r}_1,\omega)\rcb
        % \\
        &=\mathbf{G}^{(0)}(\mathbf{R}_2,\mathbf{R}_2,\omega)+\mathbf{G}^{(0)}(\mathbf{R}_2,\mathbf{R}_1,\omega)\cdot\boldsymbol{\alpha}_1(\omega)\cdot\lsb\mathbf{I}-\mathbf{G}^{(0)}(\mathbf{R}_1,\mathbf{R}_1,\omega)\cdot\boldsymbol{\alpha}_1(\omega)\rsb^{-1}\cdot\mathbf{G}^{(0)}(\mathbf{R}_1,\mathbf{R}_2,\omega),
        % \\
        % &\equiv \mathbf{G}^{(0)}(\mathbf{r}_2,\mathbf{r}_2,\omega)+\mathbf{G}^{(0)}(\mathbf{r}_2,\mathbf{r}_1,\omega)\cdot\boldsymbol{\alpha}_1(\omega)\cdot\widetilde{\mathbf{G}}_{\mathbf{r}_2,\mathbf{r}_1}(\omega),
    \end{split}
\end{equation}
and also $\mathbf{G}^{(0)}(\mathbf{r},\mathbf{r}',\omega)\equiv\mathbf{G}^{\rm B}(\mathbf{r},\mathbf{r}',\omega)$.
One can put all these together, now that we have all of the contributions only in terms of the background Green's tensor, and obtain a unified formula for the total two-body scattering Green's tensor in terms of the background Green's tensor and the dipoles' polarizabilities only.

The dipoles of interest have a
preferred polarization, such as $s$-polarized in the main text,
where 
$\boldsymbol{\alpha}_n(\omega)=\hat{\mathbf{e}}_n\,{\alpha}_n(\omega)\,
\hat{\mathbf{e}}_n$.
%However, this procedure can be further simplified for most problems assuming that the dipoles are isotropic characterized with the scalar polarizability, $\boldsymbol{\alpha}_n(\omega)={\alpha}_n(\omega)\mathbf{I}$. 
Hence, the desired Green's tensor reads~\cite{Kristensen_Decay_2011}
\begin{equation}
    \begin{split}
        \mathbf{G}^{(2)}(\mathbf{R}_1,\mathbf{R}_2,\omega)&=\frac{\widetilde{\mathbf{G}}_{\mathbf{R}_1,\mathbf{R}_2}(\omega)+\widetilde{\mathbf{G}}_{\mathbf{R}_1,\mathbf{R}_1}(\omega)\,\alpha_1(\omega)\,\widetilde{G}_{R_1,R_2}(\omega)}{1-\widetilde{G}_{R_2,R_1}(\omega)\,\alpha_1(\omega)\,\widetilde{G}_{R_1,R_2}(\omega)\,\alpha_2(\omega)},
    \end{split}
\end{equation}
where 
% \begin{equation}
% {\bf G}^{(2)}({\bf r}_1, {\bf r}_2,\omega)
% =\lsb{\mathbf{I}- \widetilde{\mathbf{G}}_{\mathbf{r}_2,\mathbf{r}_1}(\omega)\cdot\boldsymbol{\alpha}_1(\omega)\cdot \widetilde{\mathbf{G}}_{\mathbf{r}_{1},\mathbf{r}_2}(\omega) \cdot \boldsymbol{\alpha}_2(\omega)}
% \rsb^{-1}\cdot\lsb{\widetilde{\mathbf{G}}_{\mathbf{r}_1,\mathbf{r}_2}(\omega) +
% \widetilde{\mathbf{G}}_{\mathbf{r}_1,\mathbf{r}_1}(\omega) \cdot\boldsymbol{\alpha}_1(\omega) \cdot\widetilde{\mathbf{G}}_{\mathbf{r}_1,\mathbf{r}_2}(\omega)}\rsb
% ,
% \end{equation}
% \begin{equation}
% {\bf G}^{(2)}({\bf r}_1, {\bf r}_2,\omega)
% =
% \lsb{\mathbf{I}- \widetilde{\mathbf{G}}_{\mathbf{r}_2,\mathbf{r}_1}(\omega)\cdot\boldsymbol{\alpha}_1(\omega)\cdot \widetilde{\mathbf{G}}_{\mathbf{r}_{\red 2?},\mathbf{r}_2}(\omega) \cdot\boldsymbol{\alpha}_2(\omega)}\rsb^{-1}\cdot\lsb{\widetilde{\mathbf{G}}_{\mathbf{r}_1,\mathbf{r}_2}(\omega) +
% \widetilde{\mathbf{G}}_{\mathbf{r}_1,\mathbf{r}_1}(\omega) \cdot\boldsymbol{\alpha}_1(\omega) \cdot\widetilde{\mathbf{G}}_{\mathbf{r}_1,\mathbf{r}_2}(\omega)}\rsb,
% \end{equation}
\begin{equation}
\widetilde{\mathbf{G}}_{\mathbf{R}_n,\mathbf{R}_{n'}}(\omega)
= 
\frac{{\bf G}^{\rm B}({\bf R}_n,{\bf R}_{n'},\omega)}{1 - \widetilde{G}_{R_n,R_{n'}}(\omega) {\alpha}_{n'}(\omega) },
% = 
% \widetilde{\mathbf{G}}_{\mathbf{r}_b,\mathbf{r}_a}(\omega)
% =
% \lsb{\mathbf{I} - {\bf G}^{\rm B}({\bf r}_a,{\bf r}_a,\omega)\cdot\boldsymbol{\alpha}_a(\omega) }\rsb^{-1}\cdot{{\bf G}^{\rm B}({\bf r}_b,{\bf r}_a,\omega)}
% .
\label{eq:Gdd}
\end{equation}
where $\widetilde{G}_{R_n,R_{n'}}(\omega)\equiv\hat{\mathbf{e}}_n\cdot\widetilde{\mathbf{G}}_{\mathbf{R}_n,\mathbf{R}_{n'}}(\omega)\cdot\hat{\mathbf{e}}_{n'}$, and 
\begin{equation}
\alpha_n(\omega) = \frac{d_n^2 2 \omega_n}{\epsilon_0 \hbar(\omega_n^2-\omega^2)},
\end{equation}
is the semiclassical formula for the polarizability (units of volume) of a LO when using the dipole definition $d_n$;
%($\hbar$ is needed for the units consistency and is also consistent with the PyCharge formalism, where 
the connection between this polarizibility model and the classical mass model is obtained via $q_n^2/m_{{\rm eff},n} = 2 \omega_n d_n^2/\hbar$, as also mentioned in the main text.
Note that in the limit of neglecting the dipoles' self-interaction (small Lamb shift and 
background radiative decay), then $\widetilde{\mathbf{G}}_{\mathbf{R}_n,\mathbf{R}_{n'}}(\omega)={{\bf G}^{\rm B}({\bf R}_n,{\bf R}_{n'},\omega)}$.
% Also, here, the dipoles are in free space [$\epsilon(\mathbf{r})=1$] so that $\mathbf{G}^{\rm B}({\bf R}_n,{\bf R}_{n'},\omega)=\mathbf{G}^{(0)}({\bf R}_n,{\bf R}_{n'},\omega)\equiv \mathbf{G}_0({\bf R}_n,{\bf R}_{n'},\omega)$, the free space Green's function.
It is now easy to obtain the spectral resonances
for any medium described through
${\bf G}^{\rm B}$.

For a
 homogeneous background system, 
the Green function is known 
analytically~\cite{Scheel_Spontaneous_1999}:
\begin{equation}
\begin{split}
    {\mathbf{G}}^{\mathrm{\rm B}}(\mathbf{r},\mathbf{r}',\omega)&
=  - \frac{\delta(\mathbf{r}-\mathbf{r}')}{3 n^2_{\rm B}}{\mathbf{I}}
\\
&\hspace{0.5cm}+  \frac{{ k_{0}^{2}} \exp \left(\ii k_{\rm B} \lvert \mathbf{r}-\mathbf{r}'\rvert\right)}{4 \pi \lvert\mathbf{r}-\mathbf{r}'\rvert} \lsb\left(1+\frac{\ii k_{\rm B} \lvert\mathbf{r}-\mathbf{r}'\rvert -1}{k^{2}_{\rm B} \lvert\mathbf{r}-\mathbf{r}'\rvert^{2}}\right) {\mathbf{I}}  + \left(\frac{3-3 \ii k_{\rm B} \lvert\mathbf{r}-\mathbf{r}'\rvert-k^{2}_{\rm B} \lvert\mathbf{r}-\mathbf{r}'\rvert^{2}}{k^{2}_{\rm B} \lvert\mathbf{r}-\mathbf{r}'\rvert^{2}}\right)\frac{\lp\mathbf{r}-\mathbf{r}'\rp\lp\mathbf{r}-\mathbf{r}'\rp}{\lvert\mathbf{r}-\mathbf{r}'\rvert^2}\rsb,
\end{split}
    \label{eq:G_transverse}
\end{equation} 
with $k_0=\omega/c$ and $k_{\rm B}=n_{\rm B}k_0$.
Interestingly, we can regard our desired near-field coupling term as either coming from the total fields, transverse and longitudinal, or in terms of only the longitudinal fields. 
This is because
${\mathbf{G}}^{\mathrm{\rm B}}(\mathbf{r},\mathbf{r}',\omega) = {\mathbf{G}}^{\mathrm{\rm B}}_T(\mathbf{r},\mathbf{r}',\omega) + {\mathbf{G}}^{\mathrm{\rm B}}_L(\mathbf{r},\mathbf{r}',\omega)$,
where 
\begin{equation}
\begin{split}
{\mathbf{G}}_T^{\mathrm{\rm B}}(\mathbf{r},\mathbf{r}',\omega) & = 
  \frac{\mathbf{I} - \lp\mathbf{r}-\mathbf{r}'\rp\lp\mathbf{r}-\mathbf{r}'\rp}{4\pi \lvert\mathbf{r}-\mathbf{r}'\rvert^5}
  \\
  &\hspace{0.5cm}+
  \frac{{ k_{0}^{2}} \exp \left(\ii k_{\rm B} \lvert\mathbf{r}-\mathbf{r}'\rvert\right)}{4 \pi \lvert\mathbf{r}-\mathbf{r}'\rvert} \lsb\left(1+\frac{\ii k_{\rm B} \lvert\mathbf{r}-\mathbf{r}'\rvert- 1}{k^{2}_{\rm B} \lvert\mathbf{r}-\mathbf{r}'\rvert^{2}}\right) {\mathbf{I}}  + \left(\frac{3-3 \ii k_{\rm B} \lvert\mathbf{r}-\mathbf{r}'\rvert-k^{2}_{\rm B} \lvert\mathbf{r}-\mathbf{r}'\rvert^{2}}{k^{2}_{\rm B} \lvert\mathbf{r}-\mathbf{r}'\rvert^{2}}\right)\frac{\lp\mathbf{r}-\mathbf{r}'\rp\lp\mathbf{r}-\mathbf{r}'\rp}{\lvert\mathbf{r}-\mathbf{r}'\rvert^2}\rsb,
  \end{split}
\end{equation}
and
\begin{equation}
{\mathbf{G}}_L^{\mathrm{\rm B}}(\mathbf{r},\mathbf{r}',\omega) = 
  -\frac{\delta(\mathbf{r}-\mathbf{r}')}{3 n^2_{\rm B}}{\mathbf{I}}-\frac{1}{4\pi \lvert\mathbf{r}-\mathbf{r}'\rvert^3}\lsb 
  \mathbf{I} -\frac{ \lp\mathbf{r}-\mathbf{r}'\rp\lp\mathbf{r}-\mathbf{r}'\rp}{\lvert\mathbf{r}-\mathbf{r}'\rvert^2}\rsb.
\end{equation}
We highlight that these expressions include a sum over all modes and the 
two Green's tensors satisfy a fundamental completeness relation,  ${\bf G}_T({\bf r},{\bf r}')+{\bf G}_L({\bf r},{\bf r}') = \delta({\bf r}-{\bf r}') {\bf I}$. 

Next, for the interaction between the dipoles one can simply employ the dyadic Green's function approach and write the interaction energy between the two dipole via the well-known relation of~\cite{Cohen-Tannoudji_Photons_1997,Jackson_Classical_1999,Kristensen_Decay_2011} $H_{dd}=-\mathbf{d}_1\cdot\mathbf{E}(\mathbf{R}_1,\omega)=-\mathbf{d}_2\cdot\mathbf{E}(\mathbf{R}_2,\omega)$, with $\mathbf{E}(\mathbf{R}_n,\omega)={\mathbf{G}}(\mathbf{R}_n,\mathbf{R}_{m\neq n},\omega)\cdot \mathbf{d}_{m\neq n}/\epsilon_0$, as
\begin{equation}
    \begin{split}
        H_{dd}= -\frac{\mathbf{d}_1\cdot{\mathbf{G}}(\mathbf{R}_1,\mathbf{R}_2,\omega)\cdot\mathbf{d}_2}{\epsilon_0},
    \end{split}
    \label{H_dd_GF}
\end{equation}
in agreement with Eq.~\eqref{g_ij} of the main text, such that $H_{dd}=\hbar g_{12}(\beta_1+\beta^*_1)(\beta_{2}+\beta^*_{2})$.
For our interests here, we can neglect radiative loss 
and also considering two dipoles in close proximity,
in free space [$\epsilon(\mathbf{r})=1$];
then for two $s$-polarized dipoles, we include the near-field contributions (only $R^{-3}$
to the Green's function), so that
\begin{equation}
{\mathbf{G}}(\mathbf{R}_1,\mathbf{R}_2,\omega)\approx {\bf G}^{\rm B}(\mathbf{R}_1,\mathbf{R}_2,\omega)
\approx -\frac{e^{i k_0 R}}
{4 \pi\epsilon_0 R^3} {\bf I} \approx 
-\frac{1}
{4 \pi\epsilon_0 R^3} {\bf I},
\end{equation}
where $R=\lVert{\bf R}\rVert= \lVert{\bf R}_1 - {\bf R}_2\rVert$.
Therefore, the classical Hamiltonian simply becomes $H=\sum_{n=1,2}H_{d,n}+H_{dd}$, where the interaction term $H_{dd}$, as denoted above, is only characterized by the position-position interaction via the Green's tensor. 

%Indeed, this furthur yields, by letting $\mathbf{d}_n=q_nq_{d,n}=q_n\sqrt{\hbar/(2\omega_{n}m_{{\rm eff},n})}\,(\beta_n+\beta_n^*)$.
% , the same Hamiltonian as in Eq.~\eqref{HDipDip_PositionMomentum} in the dipoles' conjugate position-momentum representation, or in Eq~\eqref{HDipDip} in the dipoles' normal mode complex amplitude representation.
% Moreover, clearly, one obtains the same classical EoM given in Eq.~\eqref{MatrixHcl_EoM}, and with identical dipoles' parameters in Eq.~\eqref{MatrixH_EoM}.
The normal mode frequencies for the hybrid system is then calculated via the roots of the determinant of the classical dynamical matrix for the 
equations of motion, yielding the solution
\begin{equation}
    \begin{split}
        \omega_{\pm}^2={\frac{\omega_{1}^2+\omega_{2}^2}{2}\pm\frac{1}{2}\sqrt{(\omega_{1}^2-\omega_{2}^2)^2+16g^2\omega_{1}\omega_{2}}},
    \end{split}
\end{equation}
which simplifies to 
$\omega_{\pm}=\sqrt{\omega_0^2\pm 2g\omega_0}$
when 
$\omega_{1}=\omega_{2}=\omega_0$, as  in the 
main text. 
% Note one might consider instability of the lower mode as it becomes imaginary for the range of $g>\omega_d$. However, this is not an issue because the validity of the dipole approximation implies $R>4a$ so that the coupling factor never reaches to those high values.
% \sh{no point to repeat this, either here or in main text, but probably main text is better}

Lastly, the second quantization of the dipole-dipole Hamiltonian can simply be done by promoting the normal-mode complex amplitudes to operators via $\beta_n\to b_n$ and $\beta_n^*\to b_n^\dagger$ to write
 \begin{equation}
    H
       =\hbar \omega_{0}  b_1^\dagger b_1
       +\hbar \omega_{0}  b_2^\dagger b_2
       +\hbar g(b_1+b_1^\dagger)(b_2+b_2^\dagger),
    \label{HQED_dip-dip}
\end{equation}
precisely coinciding with 
Eq.~\eqref{eq:dd3}
and 
 Eq.~\eqref{H_qm} of the main text.
As mentioned also in the main text, this
form  is identical to a Hopfield-like model Hamiltonian without the diamagnetic term~\cite{Hopfield_Theory_1958,DeLiberato_Light-Matter_2014,Qingtian_Polaritonic_2023}, and is justified here 
rigorously from Maxwell's equations
(which of course are used to construct the Green's function solutions above).

\end{document}